\documentclass[global,twocolumn]{svjour2}

\usepackage{graphicx}
\usepackage{amsmath}
\usepackage{amssymb}
\usepackage{bm}
\usepackage{psfrag}

\journalname{Applied Physics B}

\begin{document}

\title{Born expansion of the Casimir-Polder interaction of a
ground-state atom with dielectric bodies}

\author{Stefan Yoshi Buhmann\thanks{\emph{Electronic address:}
 \texttt{s.buhmann@tpi.uni-jena.de}} 
 \and Dirk-Gunnar Welsch
}

\institute{Theoretisch-Physikalisches Institut,
Friedrich-Schiller-Universit\"{a}t Jena,
Max-Wien-Platz 1, 07743 Jena, Germany,
Phone: +49-3641-947100, Fax: +49-3641-947102}

\date{Received: date / Revised version: date}

\maketitle


\begin{abstract}
Within leading-order perturbation theory, the Casimir-Polder potential
of a ground-state atom pla\-ced within an arbitrary arrangement of
dispersing and absorbing linear bodies can be expressed in terms of
the polarizability of the atom and the scattering Green tensor of
the body-assisted electromagnetic field. Based on a Born series of the
Green tensor, a systematic expansion of the Casimir-Polder potential
in powers of the electric susceptibilities of the bodies is presented.
The Born expansion is used to show how and under which conditions the
Casimir-Polder force can be related to microscopic many-atom
van der Waals forces, for which general expressions are presented. 
As an application, the Casimir-Polder potentials of an atom near a
dielectric ring and an inhomogeneous dielectric half space are studied
and explicit expressions are presented that are valid up to second
order in the susceptibility.

\noindent
\textbf{PACS}
12.20.-m  Quantum electrodynamics --
34.50.Dy  Interactions of atoms and molecules with surfaces; photon 
          and electron emission; neutralization of ions --
34.20.-b  Interatomic and intermolecular potentials and for\-ces,
          potential energy surfaces for collisions --
42.50.Nn  Quantum optical phenomena in absorbing, dispersive and
          conducting media
\end{abstract}


\section{Introduction}
\label{sec0}

The forces of electromagnetic origin that arise between electrically
neutral, unpolarized but polarizable objects are commonly known as
dispersion forces
\cite{Dzyaloshinskii61,Langbein74,Mahanty76,Hinds91,Milonni94}. 
They were first addressed within the context of quantum
electrodynamics (QED) by Casimir and Polder
\cite{Casimir48,Casimir48b}, who showed that they may be attributed to
the vacuum fluctuations of the electromagnetic field. In accordance
with the different nature of the interacting objects, one may
distinguish between three types of dispersion forces, na\-mely the
forces between atoms---in the following referred to as van der Waals
(vdW) forces, the forces between atoms and macroscopic bodies---in the
following referred to as Casimir-Polder (CP) for\-ces, and the forces
between macroscopic bodies---in the following referred to as Casimir
forces.
   
Dispersion forces play a major role in the understanding of many
phenomena, and they can be a useful or disturbing factor in modern
applications. Apart from being crucial for the understanding of many
structures and processes in biochemistry \cite{Nelson02}, they are
responsible for the remarkable climbing skills of some gecko
\cite{Autumn02} and spider species \cite{Kesel04}; the construction of
atomic-force microscopes is essentially based on dispersion forces
\cite{Binnig86}, while they are also responsible for the problem of
sticking in nanotechnology \cite{Henkel04}. In particular, CP forces
between atoms and macroscopic bodies are needed for an understanding
of the adsorption of atoms and molecules to surfaces \cite{Bruch83};
they can be used in atom optics to construct atomic mirrors
\cite{Shimizu02}, while they have also been found to severely
limit the lifetime of atoms stored on atom chips \cite{Lin04}.

The study of CP forces which were first predicted for the idealized
situation of a ground-state atom interacting with a perfectly
conducting plate \cite{Casimir48} has since been greatly extended.
Various planar geometries like the semi-infinite half space
\cite{McLachlan63,McLachlan63b,Tiko93,Enderlein99}, plates of finite
thickness \cite{Zhou95}, two-layered plates \cite{Wylie84} or planar
cavities \cite{Zhou95,Jhe91} have been treated, the most general
planar geometry being the planar multilayer system with an arbitrary
number of layers \cite{Buhmann05,Buhmann05b}. Systems with spherical
\cite{Marvin82,Girard89} or cylindrical symmetries
\cite{Marvin82,Boustimi02} have also been considered. It
should be mentioned that some theoretical approaches (in particular,
those based on normal-mode quantization, e.g., 
Refs.~\cite{Casimir48,Tiko93,Enderlein99,Zhou95,Jhe91,Marvin82})
require a separate treatment for each specific geometry, whereas
others (in particular, the methods based on linear response theory,
e.g.,
Refs.~\cite{McLachlan63,McLachlan63b,Wylie84,Girard89,Boustimi02})
lead to general expressions that are geometry-independent. 

Recently, the problem has been studied within the frame of
macroscopic QED in dispersing and absorbing media and an exact
derivation of a general expression for the CP force has been given
\cite{Buhmann05,Buhmann04,Buhmann04b}. Although the problem of
calculating CP forces (or equivalently, the respective CP potentials)
is thus formally solved, explicit evaluation requires knowledge of the
(classical) Green tensor for the body-assisted electromagnetic field,
which is (analytically) available only for a very limited class of
geometries. In particular, inhomogeneous bodies or bodies of exotic
shapes have not yet been treated. Nevertheless, it shall be
demonstrated in this paper that the general solution---in combination
with a Born expansion of the Green tensor---may serve as a starting
point for the systematic study of a wide class of geometries.

Furthermore, the Born expansion helps making general statements
about two fundamental issues regarding CP forces. First, it may answer
the question of whether and to what extent CP forces are additive.
Second, it can be used to clarify the microscopic origin of CP
forces. It is known that up to linear order in the electric
susceptibility the force between an atom and a macroscopic body that
may be regarded composed of atom-like constituents can be obtained by
summation of two-atom (microscopic) vdW forces
\cite{Milonni94,Lifshitz56}; an analogous relation between the Casimir
force and CP forces can be established
\cite{Lifshitz56,Parsegian74,Schwinger78,Mil92,Kup92,Barton01,Raa05}.
It is
also known that pairwise summation fails at higher order in the
susceptibility \cite{Renne67,Milonni92b}, where many-atom interactions
begin to play a role \cite{Axilrod43,Aub60,Power85,Power94,Cirone96};
in fact, it has been shown that an infinite series of many-atom
interactions must be included in order to derive the CP force between
an atom and a semi-infinite dielectric half space microscopically 
\cite{Renne67}. 

The article is organized as follows. In Sec.~\ref{sec1} the Born
expansion of the CP potential of an atom pla\-ced within an arbitrary
arrangement of locally, linearly, and causal\-ly responding isotropic
dielectric bodies is given. The results are then used to elucidate
the relation to microscopic descriptions (Sec.~\ref{sec3}), and to
study some specific geometries (Sec.~\ref{sec2}). Finally, a summary
is given in Sec.~\ref{sec4}.


\section{Born expansion}
\label{sec1}

Consider a neutral, non-polar, ground-state atomic system $A$
such as an atom or a molecule (briefly referred to as atom in the
following) at position $\vec{r}_A$ which is placed in a free-space
region within an arbitrary arrangement of linear dielectric bodies.
The system of bodies is characterized by the (relative) permittivity
$\varepsilon(\vec{r},\omega)$, which is a spatially varying,
complex-valued function of frequency, with the Kramers-Kronig
relations being satisfied. Within leading-or\-der perturbation theory,
the CP force on the atom due to the presence of the bodies can be
derived from the CP potential (see, e.g., Ref.~\cite{Buhmann04})
\begin{equation}
\label{eq1}
U_A(\vec{r}_A)
= \frac{\hbar\mu_0}{2\pi}
 \int_0^{\infty} \mathrm{d} u \,u^2 \alpha_A(iu)
 \,\mathrm{Tr}\,
 \tens{G}^{(1)}(\vec{r}_A,\vec{r}_A,iu)
\end{equation}
according to
\begin{equation}
\label{eq2}
\vec{F}_A(\vec{r}_A)
=-\bm{\nabla}_{\!\!A}U_A(\vec{r}_A)
\end{equation}
($\bm{\nabla}_{\!\!A}$ $\!\equiv$
$\!\bm{\nabla}_{\!\vec{r}_A}$). In Eq.~(\ref{eq1}),
\begin{equation}
\label{eq3}
\alpha_A(\omega)=\lim_{\epsilon\to 0}\frac{2}{3\hbar}\sum_n
 \frac{\omega_{n0}^A|\vec{d}^A_{0n}|^2}
 {(\omega_{n0}^A)^2-\omega^2-i\omega\epsilon}
\end{equation}
is the ground-state polarizability of the atom in lowest order of
perturbation theory [$\omega_{n0}^A$ $\!\equiv$ $\!(E_n^A$ $\!-$
$\!E_0^A)/\hbar$, (bare) atomic transition frequencies;
$\vec{d}^A_{0n}$ $\!\equiv$ $\!\langle 0_A|
\hat{\vec{d}}_A|n_A\rangle$, electric-dipole transition matrix
elements of the atom], and $\tens{G}^{(1)}(\vec{r},\vec{r}',iu)$
is the scattering part of the classical Green tensor of the
body-assisted electromagnetic field,
\begin{equation}
\label{eq4}
\tens{G}(\vec{r},\vec{r}',\omega)
=\tens{G}^{(0)}(\vec{r},\vec{r}',\omega)
 +\tens{G}^{(1)}(\vec{r},\vec{r}',\omega)
\end{equation}
[$\tens{G}^{(0)}(\vec{r},\vec{r}',\omega)$, vacuum part], which
is the solution to the equation
\begin{equation}
\label{eq5}
\biggl[\bm{\nabla}\times\bm{\nabla}\times
 -\frac{\omega^2}{c^2}\,\varepsilon(\vec{r},\omega)\biggr]
 \tens{G}(\vec{r},\vec{r}',\omega)
=\delta(\vec{r}-\vec{r}')\tens{I}
\end{equation}
($\tens{I}$, unit tensor) together with the boundary condition
\begin{equation}
\label{eq6}
\tens{G}(\vec{r},\vec{r}',\omega)\to 0
 \quad\mbox{for}\quad|\vec{r}-\vec{r}'|\to\infty.
\end{equation}

Suppose now that
\begin{equation}
\label{eq7}
\varepsilon(\vec{r},\omega)
=\overline{\varepsilon}(\vec{r},\omega)
 +\chi(\vec{r},\omega) ,
\end{equation}
with the Green tensor $\overline{\tens{G}}(\vec{r},\vec{r}',\omega)$,
which is the solution to
\begin{equation}
\label{eq7.1}
\biggl[\bm{\nabla}\times\bm{\nabla}\times
 -\frac{\omega^2}{c^2}\,\overline{\varepsilon}(\vec{r},\omega)\biggr]
 \overline{\tens{G}}(\vec{r},\vec{r}',\omega)
=\delta(\vec{r}-\vec{r}')\tens{I},
\end{equation}
being known. A (formal) solution to Eq.~(\ref{eq5}) can then be given
by the Born series
\begin{align}
\label{eq8}
&\tens{G}(\vec{r},\vec{r}',\omega)
 =\overline{\tens{G}}(\vec{r},\vec{r}',\omega)\nonumber\\
&\hspace{1ex}+\sum_{k=1}^\infty\Bigl(\frac{\omega}{c}\Bigr)^{2k}
 \Biggl[\prod_{j=1}^k\int\mathrm{d}^3s_j\,
 \chi(\vec{s}_j,\omega)\Biggr]\nonumber\\
&\quad\times
\overline{\tens{G}}(\vec{r},\vec{s}_1,\omega)\cdot
\overline{\tens{G}}(\vec{s}_1,\vec{s}_2,\omega)\cdots
\overline{\tens{G}}(\vec{s}_k,\vec{r}',\omega),
\end{align}
as can be easily verified using Eq.~(\ref{eq7.1}) together with
\begin{align}
\label{eq8.1}
&\biggl[\bm{\nabla}\!\times\!\bm{\nabla}\!\times\!
 -\frac{\omega^2}{c^2}\,\overline{\varepsilon}(\vec{r},\omega)\biggr]
 \sum_{k=1}^\infty\Bigl(\frac{\omega}{c}\Bigr)^{2k}
 \Biggl[\prod_{j=1}^k\int\mathrm{d}^3s_j\,
 \chi(\vec{s}_j,\omega)\Biggr]\nonumber\\
&\quad\times
 \overline{\tens{G}}(\vec{r},\vec{s}_1,\omega)\cdot
 \overline{\tens{G}}(\vec{s}_1,\vec{s}_2,\omega)\cdots
 \overline{\tens{G}}(\vec{s}_k,\vec{r}',\omega)\nonumber\\
&\qquad=\Bigl(\frac{\omega}{c}\Bigr)^2\chi(\vec{r},\omega)
 \tens{G}(\vec{r},\vec{r}',\omega).
\end{align}
Combining Eqs.~(\ref{eq1}), (\ref{eq4}), and (\ref{eq8}), we find that
the CP potential can be expanded as
\begin{equation}
\label{eq9.1}
U_A(\vec{r}_A)
=\overline{U}_A(\vec{r}_A)
+\sum_{k=1}^\infty\Delta_kU_A(\vec{r}_A),
\end{equation}
where
\begin{equation}
\label{eq9.2}
\overline{U}_A(\vec{r}_A)
=\frac{\hbar\mu_0}{2\pi}
 \int_0^{\infty} \mathrm{d} u \,u^2 \alpha_A(iu)
 \,\mathrm{Tr}\,\overline{\tens{G}}^{(1)}
 (\vec{r}_A,\vec{r}_A,iu)
\end{equation}
is the CP potential due to $\overline{\varepsilon}(\vec{r},\omega)$,
and
\begin{align}
\label{eq9}
&\Delta_kU_A(\vec{r}_A)
 = \frac{(-1)^k\hbar\mu_0}{2\pi c^{2k}}\nonumber\\
&\hspace{1ex}\times
 \int_0^{\infty} \mathrm{d} u \,u^{2k+2}
 \alpha_A(iu)
 \Biggl[\prod_{j=1}^k\int\mathrm{d}^3s_j\,
 \chi(\vec{s}_j,iu)\Biggr]
 \nonumber\\
&\hspace{1ex}\times\mathrm{Tr}\bigl[
 \overline{\tens{G}}(\vec{r}_A,\vec{s}_1,iu)\cdot
\overline{\tens{G}}(\vec{s}_1,\vec{s}_2,iu)\cdots
\overline{\tens{G}}(\vec{s}_k,\vec{r}_A,iu)\bigr]
\end{align}
is the contribution to the potential that is of $k$th order in
$\chi(\vec{r},\omega)$. The Born expansion of the CP potential as
given by Eqs.~(\ref{eq9.1})--(\ref{eq9}) can be used to
(systematically) calculate the potential in scenarios where a basic
arrangement of bodies for which the Green tensor is known is (weakly)
disturbed, e.g., by additional bodies or inhomogeneities such as
surface roughness.

Let us apply  Eq.~(\ref{eq9.1})--(\ref{eq9}) to the case of
arbitrarily shaped, weakly dielectric bodies, so that we may let
$\overline{\varepsilon}(\mathbf{r},\omega)$ $\!\equiv$ $\!1$, and
hence
\begin{equation}
\label{eq14}
U_A(\vec{r}_A)
 =\sum_{k=1}^\infty\Delta_kU_A(\vec{r}_A),
\end{equation}
where $\Delta_kU_A(\vec{r}_A)$ is given by Eq.~(\ref{eq9}), with
\begin{align}
\label{eq15}
\overline{\tens{G}}(\vec{r},\vec{r}',iu)
&=\tens{G}_\mathrm{V}(\vec{r},\vec{r}',iu)\nonumber\\
&=\frac{1}{4\pi}\biggl[\tens{I}-\Bigl(\frac{c}{u}\Bigr)^2
 \bm{\nabla}\otimes\bm{\nabla}\biggr]\frac{e^{-\frac{u\rho}{c}}}{\rho}
 \nonumber\\
&=\frac{1}{3}\Bigl(\frac{c}{u}\Bigr)^2\delta(\bm{\rho})\tens{I}
 +\tens{H}_\mathrm{V}(\vec{r},\vec{r}',iu)
\end{align}
being the vacuum Green tensor (see, e.g., Ref.~\cite{Knoll01}), where
\begin{equation}
\label{eq16}
\tens{H}_\mathrm{V}(\vec{r},\vec{r}',iu) =
 \frac{c^2e^{-\frac{u\rho}{c}}}{4\pi u^2\rho^3}
 \biggl[a\Bigl(\frac{u\rho}{c}\Bigr)\tens{I}
 -b\Bigl(\frac{u\rho}{c}\Bigr)
 \hat{\bm{\rho}}\otimes\hat{\bm{\rho}}\biggr],
\end{equation}
\begin{align}
\label{eq16.1}
a(x)&=1+x+x^2,\\
\label{eq16.2}
b(x)&=3+3x+x^2
\end{align}
($\bm{\rho}$ $\!\equiv$
$\!\vec{r}-\vec{r}'$; $\rho$ $\!\equiv$ $\!|\bm{\rho}|$;
$\hat{\bm{\rho}}$ $\!\equiv$ $\!\bm{\rho}/\rho$). Combining
Eqs.~(\ref{eq9}) and (\ref{eq15}) [together with
Eqs.~(\ref{eq16})--(\ref{eq16.2})], one easily finds that to linear
order in $\chi$ the CP potential reads
\begin{align}
\label{eq17}
U_A(\vec{r}_A) &=
\Delta_1U_A(\vec{r}_A)
= -\frac{\hbar}{32\pi^3\varepsilon_0}
 \int_0^{\infty} \mathrm{d} u\,
 \alpha_A(iu)\nonumber\\
&\quad\times\int\mathrm{d}^3s\,\chi(\vec{s},iu)
 \,\frac{g_2(u|\vec{r}_A-\vec{s}|/c)}
 {|\vec{r}_A-\vec{s}|^6},
\end{align}
where
\begin{equation}
\label{eq18}
g_2(x)=2e^{-2x}(3+6x+5x^2+2x^3+x^4).
\end{equation}
In this approximation the CP force is simply a volume integral over
attractive central forces, as is seen from
\begin{align}
\label{eq18.1}
&\bm{\nabla}\biggl[\frac{g_2(ur/c)}{r^6}\biggr]
 =-\frac{\hat{\vec{r}}}{r^7}\bigl[
 6g_2(ur/c)-(ur/c)g^\prime_2(ur/c)\bigr]\nonumber\\
&\quad=-\frac{4\hat{\vec{r}}}{r^7}
 \bigl[e^{-2x}(9\!+\!18x\!+\!16x^2\!+\!8x^3\nonumber\\
&\quad\hspace{8ex}+\!3x^4\!+\!x^5)\bigr]_{x=ur/c}
\end{align}
($r$ $\!\equiv$ $\!|\vec{r}|$, $\hat{\vec{r}}$ $\!\equiv$
$\!\vec{r}/r$).

In the retarded (long-distance) limit, i.e.,
\begin{align}
\label{eq19}
r_-\gg
 \frac{c}{\omega_A^-}\,,\quad
r_-\gg \frac{c}{\omega_\mathrm{M}^-}\,,
\end{align}
where $r_-$ $\!\equiv$
$\min\{|\vec{r}_A-\vec{s}|:\chi(\vec{s})\neq 0\}$ is the minimum
distance of the atom to any of the bodies, $\omega_A^-$ $\!\equiv$
$\!\mathrm{min}\{\omega_{n0}^A|n$ $\!=$ $\!1,2,\ldots\}$ is
the lowest atomic transition frequency, and $\omega_\mathrm{M}^-$ is
the lowest resonance frequency of the dielectric material,
the exponential factor in $g_2(x)$ effectively limits the
$u$-integral in Eq.~(\ref{eq17}) to a region where
\begin{equation}
\label{eq20}
\alpha_A(iu)\simeq\alpha_A(0),\
 \chi(\vec{s},iu)\simeq\chi(\mathbf{s},0),
\end{equation}
so Eq.~(\ref{eq17}) reduces to
\begin{eqnarray}
\label{eq21}
\Delta_1U_A(\vec{r}_A)
&=&-\frac{\hbar c\alpha_A(0)}{32\pi^3\varepsilon_0}
 \int\mathrm{d}^3s\,\frac{\chi(\mathbf{s},0)}
 {|\vec{r}_A-\vec{s}|^7}
 \int_0^{\infty} \mathrm{d}x\, g_2(x)
 \nonumber\\
&=&-\frac{23\hbar c\alpha_A(0)}
 {64\pi^3\varepsilon_0}
 \int\mathrm{d}^3s\,
 \frac{\chi(\vec{s},0)}{|\vec{r}_A-\vec{s}|^7}\, .
\end{eqnarray}
In the nonretarded (short-distance) limit, i.e.,
\begin{align}
\label{eq22}
r_+\ll
 \frac{c}{\omega_A^+}\quad\mathrm{and/or}\quad
r_+\ll \frac{c}{\omega_\mathrm{M}^+}\,,
\end{align}
where $r_+$ $\!\equiv$
$\max\{|\vec{r}_A-\vec{s}|:
\chi(\vec{s})\neq 0\}$ is the maximum distance of the atom to
any body part, $\omega_A^+$ $\!\equiv$
$\!\mathrm{max}\{\omega_{n0}^A|n$ $\!=$ $\!1,2,\ldots\}$ is
the highest atomic transition frequency, and $\omega_\mathrm{M}^+$ is
the highest resonance frequency of the dielectric material, the
factors $\alpha_A(iu)$ and $\chi(\vec{s},iu)$ effectively limit the
$u$-integral in Eq.~(\ref{eq17}) to a region where $x$ $\!=$
$\!u|\vec{r}_A-\vec{s}|/c$ $\!\ll$ $\!1$, so we may set
\begin{equation}
\label{eq23}
g_2(x)\simeq g_2(0)=6,
\end{equation}
resulting in
\begin{equation}
\label{eq24}
\Delta_1U_A(\vec{r}_A)
=-\frac{3\hbar}{16\pi^3\varepsilon_0}
 \int_0^{\infty}\! \mathrm{d} u\, \alpha_A(iu)
 \int\mathrm{d}^3s\,
 \frac{\chi(\vec{s},iu)}{|\vec{r}_A-\vec{s}|^6}\, .
\end{equation}

The second-order contribution $\Delta_2U_A(\vec{r}_A)$ can be
separated into a single-point term and a two-point correlation term,
\begin{equation}
\label{eq24.1}
\Delta_2U_A(\vec{r}_A)
=\Delta_2^1U_A(\vec{r}_A)
+\Delta_2^2U_A(\vec{r}_A),
\end{equation}
as can be seen from Eqs.~(\ref{eq9}) and (\ref{eq15}) for $k$ $\!=$
$\!2$. The single-point term
\begin{align}
\label{eq24.2}
\Delta_2^1U_A(\vec{r}_A)
&=\frac{\hbar}{96\pi^3\varepsilon_0}
 \int_0^{\infty} \mathrm{d} u\,
 \alpha_A(iu)\nonumber\\
&\quad\times\int\mathrm{d}^3s\,\chi^2(\vec{s},iu)
 \,\frac{g_2(u|\vec{r}_A-\vec{s}|/c)}
 {|\vec{r}_A-\vec{s}|^6}\,,
\end{align}
which arises from the $\delta$-function in Eq.~(\ref{eq15}), differs
from the first-order contribution $\Delta_1U_A(\vec{r}_A)$ according
to
\begin{equation}
\label{eq24.3}
\chi(\vec{r},\omega)\mapsto-\frac{1}{3}\chi^2(\vec{r},\omega),
\end{equation}
hence its asymptotic retarded and nonretarded forms can be obtained by
applying the replacement (\ref{eq24.3}) to Eqs.~(\ref{eq21}) and
(\ref{eq24}), respectively.

The two-point correlation term is derived to be
\begin{align}
\label{eq24.4}
\Delta_2^2U_A(\vec{r}_A)
&=\frac{\hbar}{128\pi^4\varepsilon_0}
 \int_0^{\infty} \mathrm{d} u\,
 \alpha_A(iu)
 \int\mathrm{d}^3s_1\,\chi(\vec{s}_1,iu)\nonumber\\
&\quad\times
 \int\mathrm{d}^3s_2\,\chi(\vec{s}_2,iu)
 \,\frac{g_3(u,\bm{\alpha},\bm{\beta},\bm{\gamma})}
 {\alpha^3\beta^3\gamma^3},
\end{align}
where
\begin{align}
\label{eq24.5}
&g_3(u,\bm{\alpha},\bm{\beta},\bm{\gamma})
=e^{-u(\alpha+\beta+\gamma)/c}\biggl[
 3a\Bigl(\frac{u\alpha}{c}\Bigr)a\Bigl(\frac{u\beta}{c}\Bigr)
  a\Bigl(\frac{u\gamma}{c}\Bigr)
 \nonumber\\
&
-b\Bigl(\frac{u\alpha}{c}\Bigr)a\Bigl(\frac{u\beta}{c}\Bigr)
  a\Bigl(\frac{u\gamma}{c}\Bigr)
\!-a\Bigl(\frac{u\alpha}{c}\Bigr)b\Bigl(\frac{u\beta}{c}\Bigr)
  a\Bigl(\frac{u\gamma}{c}\Bigr)
 \nonumber\\
&
 -a\Bigl(\frac{u\alpha}{c}\Bigr)a\Bigl(\frac{u\beta}{c}\Bigr)
  b\Bigl(\frac{u\gamma}{c}\Bigr)
 \!+b\Bigl(\frac{u\alpha}{c}\Bigr)b\Bigl(\frac{u\beta}{c}\Bigr)
  a\Bigl(\frac{u\gamma}{c}\Bigr)
  (\hat{\bm{\alpha}}\!\cdot\!\hat{\bm{\beta}})^2
 \nonumber\\
&
 +a\Bigl(\frac{u\alpha}{c}\Bigr)b\Bigl(\frac{u\beta}{c}\Bigr)
  b\Bigl(\frac{u\gamma}{c}\Bigr)
  (\hat{\bm{\beta}}\!\cdot\!\hat{\bm{\gamma}})^2
 \nonumber\\
 &
 +b\Bigl(\frac{u\alpha}{c}\Bigr)a\Bigl(\frac{u\beta}{c}\Bigr)
  b\Bigl(\frac{u\gamma}{c}\Bigr)
  (\hat{\bm{\gamma}}\!\cdot\!\hat{\bm{\alpha}})^2
 \nonumber\\
&
 -b\Bigl(\frac{u\alpha}{c}\Bigr)b\Bigl(\frac{u\beta}{c}\Bigr)
  b\Bigl(\frac{u\gamma}{c}\Bigr)
 (\hat{\bm{\alpha}}\!\cdot\!\hat{\bm{\beta}})
 (\hat{\bm{\beta}}\!\cdot\!\hat{\bm{\gamma}})
 (\hat{\bm{\gamma}}\!\cdot\!\hat{\bm{\alpha}})\biggr],
\end{align}
with the abbreviations
\begin{alignat}{3}
\label{eq24.6}
&\bm{\alpha}\equiv\vec{r}_A-\vec{s}_1,\quad
 &\alpha\equiv|\bm{\alpha}|,\quad
 &\hat{\bm{\alpha}}\equiv\frac{\bm{\alpha}}{\alpha}\,,\\
\label{eq24.7}
&\bm{\beta}\equiv\vec{s}_1-\vec{s}_2,\quad
 &\beta\equiv|\bm{\beta}|,\quad
 &\hat{\bm{\beta}}\equiv\frac{\bm{\beta}}{\beta}\,,\\
\label{eq24.8}
&\bm{\gamma}\equiv\vec{s}_2-\vec{r}_A,\quad
 &\gamma\equiv|\bm{\gamma}|,\quad
 &\hat{\bm{\gamma}}\equiv\frac{\bm{\gamma}}{\gamma}
\end{alignat}
having been introduced [recall Eqs.~(\ref{eq16.1}) and
(\ref{eq16.2})]. Note that the two-point contribution to the CP force,
Eq.~(\ref{eq24.4}), is a double spatial integral, the integrand of
which can be attractive or repulsive, depending on the angles in the
triangle formed by the vectors $\bm{\alpha}$, $\bm{\beta}$, and
$\bm{\gamma}$.

In the retarded limit, where the inequalities (\ref{eq19})
\linebreak hold, the $u$-integral is again effectively limited to a
region where the approximations (\ref{eq20}) are valid, so
Eq.~(\ref{eq24.4}) reduces
to
\begin{align}
\label{eq24.9}
&\Delta_2^2U_A(\vec{r}_A)
=\frac{\hbar\alpha_A(0)}{128\pi^4\varepsilon_0}
 \int\mathrm{d}^3s_1\,\chi(\vec{s}_1,0)
 \nonumber\\
&\quad\times\int\mathrm{d}^3s_2\,
 \frac{\chi(\vec{s}_2,0)}{\alpha^3\beta^3\gamma^3}
 \int_0^{\infty} \mathrm{d} u\,
 g_3(u,\bm{\alpha},\bm{\beta},\bm{\gamma}).
\end{align}
We introduce the notation
\begin{equation}
\label{eq24.10}
\sigma_i\equiv
\alpha^i+\beta^i+\gamma^i,
\quad i=1,2,3
\end{equation}
and perform the $u$-integral with the aid of the relation
\begin{equation}
\label{eq24.11}
\int_0^{\infty} \mathrm{d} u\,\Bigl(\frac{u}{c}\Bigr)^j
 e^{-u\sigma_1/c}=\frac{j!\,c}{\sigma_1^{j+1}}\,.
\end{equation}
Exploiting the triangle formula
\begin{align}
\label{eq24.12}
T&\equiv 1
\!-\!(\hat{\bm{\alpha}}\!\cdot\!\hat{\bm{\beta}})^2
\!-\!(\hat{\bm{\beta}}\!\cdot\!\hat{\bm{\gamma}})^2
\!-\!(\hat{\bm{\gamma}}\!\cdot\!\hat{\bm{\alpha}})^2
\!+\!2(\hat{\bm{\alpha}}\!\cdot\!\hat{\bm{\beta}})
 (\hat{\bm{\beta}}\!\cdot\!\hat{\bm{\gamma}})
 (\hat{\bm{\gamma}}\!\cdot\!\hat{\bm{\alpha}})\nonumber\\
&=0
\end{align}
[which is a trivial consequence of
Eqs.~(\ref{eq24.6})--(\ref{eq24.8})] by adding the expression
\begin{align}
\label{eq24.13}
&6T+\frac{6T}{\sigma_1^6}\bigl\{
 \bigl[\alpha^5(\beta+\gamma)+\beta^5(\gamma+\alpha)
 +\gamma^5(\alpha+\beta)\bigr]\nonumber\\
&+7\bigl[\alpha^4(\beta^2+\gamma^2)+\beta^4(\gamma^2+\alpha^2)
 +\gamma^4(\alpha^2+\beta^2)\bigr]\nonumber\\
&+12(\alpha^3\beta^3+\beta^3\gamma^3+\gamma^3\alpha^3)
 +12\alpha\beta\gamma(\alpha^3+\beta^3+\gamma^3)\nonumber\\
&+52\alpha\beta\gamma\bigl[\alpha\beta(\alpha+\beta)
 +\beta\gamma(\beta+\gamma)+\gamma\alpha(\gamma+\alpha)
 \bigr]\nonumber\\
&+138\alpha^2\beta^2\gamma^2\bigr\}\,
\end{align}
to Eq.~(\ref{eq24.9}), the result may be written in the form
\begin{align}
\label{eq24.14}
&\Delta_2^2U_A(\vec{r}_A)
=\frac{\hbar c\alpha_A(0)}{32\pi^4\varepsilon_0}
 \int\mathrm{d}^3s_1\,\chi(\vec{s}_1,0)
 \int\mathrm{d}^3s_2
 \nonumber\\
&\quad\times
 \frac{\chi(\vec{s}_2,0)}
 {\alpha^3\beta^3\gamma^3(\alpha\!+\!\beta\!+\!\gamma)}
 \bigl[f_1(\alpha,\beta,\gamma)
 +f_2(\gamma,\alpha,\beta)
 (\hat{\bm{\alpha}}\!\cdot\!\hat{\bm{\beta}})^2\nonumber\\
&\quad\hspace{4ex}+f_2(\alpha,\beta,\gamma)
 (\hat{\bm{\beta}}\!\cdot\!\hat{\bm{\gamma}})^2
 +f_2(\beta,\gamma,\alpha)
 (\hat{\bm{\gamma}}\!\cdot\!\hat{\bm{\alpha}})^2\nonumber\\
&\quad\hspace{4ex}+f_3(\alpha,\beta,\gamma)
 (\hat{\bm{\alpha}}\!\cdot\!\hat{\bm{\beta}})
 (\hat{\bm{\beta}}\!\cdot\!\hat{\bm{\gamma}})
 (\hat{\bm{\gamma}}\!\cdot\!\hat{\bm{\alpha}})\bigr],
\end{align}
where
\begin{align}
\label{eq24.15}
&f_1(\alpha,\beta,\gamma)=
\nonumber\\&\quad 
9-39\frac{\sigma_2}{\sigma_1^2}+22\frac{\sigma_3}{\sigma_1^3}
 +54\frac{\sigma_2^2}{\sigma_1^4}
 -65\frac{\sigma_2\sigma_3}{\sigma_1^5}
+20\frac{\sigma_3^2}{\sigma_1^6}\,,\\
\label{eq24.16}
&f_2(\alpha,\beta,\gamma)=
 3\biggl[\frac{\alpha^2}{\sigma_1^2}
 +\frac{3\alpha^2(\beta+\gamma)}{\sigma_1^3}
\nonumber\\&\hspace{14ex} 
 +\frac{4\beta\gamma(3\alpha^2-\beta\gamma)}{\sigma_1^4}
-\frac{20\alpha\beta^2\gamma^2}{\sigma_1^5}\biggr],\\
\label{eq24.17}
&f_3(\alpha,\beta,\gamma)=
\nonumber\\&\quad
 -1-39\frac{\sigma_2}{\sigma_1^2}+17\frac{\sigma_3}{\sigma_1^3}
 +72\frac{\sigma_2^2}{\sigma_1^4}
 -75\frac{\sigma_2\sigma_3}{\sigma_1^5}
+20\frac{\sigma_3^2}{\sigma_1^6}
\end{align}
[recall Eqs.~(\ref{eq24.6})--(\ref{eq24.8}) as well as
Eq.~(\ref{eq24.10})].

In the nonretarded limit, where the inequalities (\ref{eq22})
hold, the $u$-integral in Eq.~(\ref{eq24.4}) is effectively limited to
a region where
\begin{align}
\label{eq24.18}
g_3(u,\bm{\alpha},\bm{\beta},\bm{\gamma})&\,\simeq
 g_3(0,\bm{\alpha},\bm{\beta},\bm{\gamma})\nonumber\\
&\,=3\bigl[1-3(\hat{\bm{\alpha}}\!\cdot\!\hat{\bm{\beta}})
 (\hat{\bm{\beta}}\!\cdot\!\hat{\bm{\gamma}})
 (\hat{\bm{\gamma}}\!\cdot\!\hat{\bm{\alpha}})\bigr]
\end{align}
[recall Eq.~(\ref{eq24.12}); note that $\max\{|\vec{s}_1-\vec{s}_2|:
\chi(\vec{s}_1)\neq 0, \linebreak \chi(\vec{s}_2)\neq 0\}$ $\!\le$
$\!2\max\{|\vec{r}_A-\vec{s}|: \chi(\vec{s})\neq 0\}$], so
Eq.~(\ref{eq24.4}) reduces to
\begin{align}
\label{eq24.19}
&\Delta_2^2U_A(\vec{r}_A)
=\frac{3\hbar}{128\pi^4\varepsilon_0}
 \int_0^{\infty} \mathrm{d} u\,
 \alpha_A(iu)
 \int\mathrm{d}^3s_1\,\chi(\vec{s}_1,iu)\nonumber\\
&\quad\times
 \int\mathrm{d}^3s_2\,\chi(\vec{s}_2,iu)
 \,\frac{1-3(\hat{\bm{\alpha}}\!\cdot\!\hat{\bm{\beta}})
 (\hat{\bm{\beta}}\!\cdot\!\hat{\bm{\gamma}})
 (\hat{\bm{\gamma}}\!\cdot\!\hat{\bm{\alpha}})}
 {\alpha^3\beta^3\gamma^3}
\end{align}
[recall Eq.~(\ref{eq24.6})--(\ref{eq24.8})].


\section{Relation to microscopic many-atom van der Waals forces}
\label{sec3}

In order to gain insight into the microscopic origin of the CP
potential as given by Eq.~(\ref{eq1}), let us suppose that the
susceptibility $\chi(\vec{r},\omega)$ is due to a collection of atoms
of polarizability $\alpha_B(\omega)$ and apply the well-known
Clausius-Mosotti formula (see, e.g., Ref.~\cite{Jackson99})
\begin{equation}
\label{eq28}
\chi(\vec{r},\omega)
 =\frac{\varepsilon_0^{-1}n(\vec{r})\alpha_B(\omega)}
 {1-\frac{1}{3}\varepsilon_0^{-1}n(\vec{r})\alpha_B(\omega)}\,,
\end{equation}
where $n(\vec{r})$ is the number density of the (medium)
atoms \mbox{[$n(\vec{r})$ $\!=$ $\!0$} for $\vec{r}$ $\!=$
$\!\vec{r}_A$]. Since $\chi(\vec{r},\omega)$ is the Fourier transform
of a (linear) response function, it must satisfy the condition
\begin{equation}
\label{eq28-1}
\chi(\vec{r},0)>\chi(\vec{r},iu)>0\quad\mathrm{for}\quad u > 0,	
\end{equation}
which, with respect to Eq.~(\ref{eq28}), implies that the 
inequality
\begin{equation}
\label{eq28.1.2}
{\textstyle\frac{1}{3}}
\varepsilon_0^{-1}n(\vec{r})\alpha_B(iu) < 1
\end{equation}
must hold. Substituting the susceptibility from Eq.~(\ref{eq28}) into
the Born series of the CP potential $U_A(\vec{r}_A)$ as given by
Eqs.~(\ref{eq9.1})--(\ref{eq9}), taking into account that the Green
tensor $\overline{\tens{G}}(\vec{r},\vec{r}',iu)$ can be decomposed as
\begin{equation}
\label{eq28.1}
\overline{\tens{G}}(\vec{r},\vec{r}',iu)
=\frac{1}{3}\Bigl(\frac{c}{u}\Bigr)^2\delta(\bm{\rho})\tens{I}
 +\overline{\tens{H}}(\vec{r},\vec{r}',iu)
\end{equation}
[recall Eqs.~(\ref{eq7.1}) and (\ref{eq15})], and recalling the
inequality (\ref{eq28.1.2}), it can be shown after some lengthy
calculation that the Born series can be rewritten as an expansion of
$U_A(\vec{r}_A)$ in terms of many-atom interaction potentials 
$U_{AB\ldots B}(\vec{r}_A,\vec{s}_1,\ldots,\vec{s}_l)$ (see
App.~\ref{AppC}),
\begin{align}
\label{eq28.1.1}
&U_A(\vec{r}_A)=\overline{U}_A(\vec{r}_A)\nonumber\\
&+\sum_{l=1}^\infty\frac{1}{l!}\Biggl[\prod_{j=1}^l
 \int\mathrm{d}^3s_j\,n(\vec{s}_j)\Biggr]
 U_{AB\ldots B}(\vec{r}_A,\vec{s}_1,\ldots,\vec{s}_l),
\end{align}
where 
\begin{align}
\label{eq28.11}
&U_{AB\ldots B}(\vec{r}_1,\ldots,\vec{r}_{l+1})
 \nonumber\\
&\quad=\frac{(-1)^l\hbar\mu_0^{l+1}}
 {(1+\delta_{1l})\pi}\int_0^\infty\mathrm{d}u\,u^{2l+2}
 \alpha_A(iu)\alpha_B^l(iu)\nonumber\\
&\qquad\times\mathcal{S}\mathrm{Tr}\bigl[
 \overline{\tens{H}}(\vec{r}_1,\vec{r}_2,iu)\cdots
 \overline{\tens{H}}(\vec{r}_{l+1},\vec{r}_1,iu)\bigr].
\end{align}
Here the symbol $\mathcal{S}$ introduces symmetrization with 
respect to $\vec{r}_1,\ldots,\vec{r}_{l+1}$ according to the rule  
\begin{align}
\label{eq28.5}
&\mathcal{S}\mathrm{Tr}\bigl[
 \overline{\tens{H}}(\vec{r}_1,\vec{r}_2,\omega)\cdots
 \overline{\tens{H}}(\vec{r}_j,\vec{r}_1,\omega)\bigr]\nonumber\\
&=\!\!\sum_{\pi\in \overline{P}(j)}\!\!
 \mathrm{Tr}\bigl[
 \overline{\tens{H}}
 (\vec{r}_{\pi(1)},\vec{r}_{\pi(2)},\omega)\cdots
 \overline{\tens{H}}(\vec{r}_{\pi(j)},\vec{r}_{\pi(1)},\omega)\bigr].
\end{align}
The sum in Eq.~(\ref{eq28.5}) runs over the maximal number of
\mbox{$j!/$ $\![(2$ $\!-$ $\!\delta_{2j})j]$} permutations $\pi$
$\!\in$ $\overline{P}(j)$ $\!\varsubsetneq$ $P(j)$ [$P(j)$ being the
permutation group of the numbers $1,\ldots,j$] that cannot be
obtained from one
another via (a) a cyclic permutation or (b) the reverse of a cyclic
permutation (cf. App.~\ref{AppC}). The potential $U_{AB\ldots
B}(\vec{r}_1,\ldots,\vec{r}_{l+1})$ is nothing but the (microscopic)
vdW potential describing the mutual interaction of a (test) atom $A$
at position $\vec{r}_1$ and $l$ (medium) atoms at different positions
$\vec{r}_2,\ldots\vec{r}_{l+1}$.

The very general equation (\ref{eq1}), which follows from QED in
causal media, gives the CP potential of an atom in the presence of
macroscopic dielectric bodies in terms of the atomic polarizability
and the (scattering) Green tensor of the body-assisted Maxwell field,
with the bodies being characterized by a spatially varying dielectric
susceptibility that is a complex function of frequency. Equation
(\ref{eq28.1.1}) clearly shows that when the susceptibility is of
Clausius-Mosotti type, i.e., Eq.~(\ref{eq28}) [together with the
inequality (\ref{eq28.1.2})] applies, then the CP potential is in fact
the result of a superposition of all possible microscopic many-atom
vdW potentials between the atom under consideration and the atoms
forming the bodies. Note that for the vacuum case,
$\overline{\varepsilon}(\vec{r},\omega)$ $\!\equiv$ $\!1$, a similar
conclusion has been drawn by combining normal mode quantization with
the Ewald-Oseen extinction theorem \cite{Milonni92b}. Moreover, from
the derivation given in App.~\ref{AppC} it can be seen that when
Eqs.~(\ref{eq28.1.1}) and (\ref{eq28.11}) [together with the
inequality (\ref{eq28.1.2})] hold, then the susceptibility must
necessarily have the form of Eq.~(\ref{eq28}).  

{F}rom the above it is clear that in order to establish the identity
(\ref{eq28.1.1}) to all orders in $\alpha_B$, it is crucial to employ
the exact relation (\ref{eq28}) between macroscopic susceptibility and
microscopic atomic polarizability rather than its linearized version
$\chi(\vec{r},\omega)$ $\!=$
$\!\varepsilon_0^{-1}n(\vec{r})\alpha_B(\omega)$, which is is known to
be sufficient for finding a correspondence between macroscopic and
microscopic potentials to linear order in $\chi$ (or $\alpha_B$,
respectively)
\cite{Milonni94,Lifshitz56,Parsegian74,Schwinger78,Mil92}.
It should be pointed out that in more general cases where the
susceptibility is not of the form (\ref{eq28}) the result of applying
the Born expansion cannot be disentangled into spatial integrals over
microscopic vdW potentials in the way given by Eq.~(\ref{eq28.1.1})
together with Eq.~(\ref{eq28.11}). Obviously, the basic constituents
of the bodies can no longer be approximated by well localized atoms.

According to Eq.~(\ref{eq7}), the expansion in Eq.~(\ref{eq28.1.1}) 
does not necessarily refer to all bodies. Hence from
Eq.~(\ref{eq28.11}) it follows that the many-atom vdW potential on an
arbitrary dielectric background described by
$\overline{\varepsilon}(\vec{r},\omega)$ reads
\begin{align}
\label{eq28.12}
&U_{A_1\ldots A_j}(\vec{r}_1,\ldots,\vec{r}_j)
 \nonumber\\
&\quad=\frac{(-1)^{j-1}\hbar\mu_0^j}
 {(1+\delta_{2j})\pi}\int_0^\infty\mathrm{d}u\,u^{2j}
 \alpha_{A_1}(iu)\cdots\alpha_{A_j}(iu)\nonumber\\
&\qquad\times\mathcal{S}\mathrm{Tr}\bigl[
 \overline{\tens{H}}(\vec{r}_1,\vec{r}_2,iu)\cdots
 \overline{\tens{H}}(\vec{r}_j,\vec{r}_1,iu)\bigr],
\end{align}
the derivation being unique when requiring the vdW potentials to be
fully symmetrized. In particular, for \linebreak
$j$ $\!=$ $\!2$ [where $j!/[(2\!-\delta_{2j})j]$ $\!=$ $\!1$, so 
the sum in the r.h.s of Eq.~(\ref{eq28.5}) contains only one term and 
symmetrization is not necessary], Eq.~(\ref{eq28.12}) agrees with the
result that can be found by calculating the change in the zero-point
energy of the system in leading-order perturbation theory
\cite{Safari05}. In the simplest case of vacuum background, i.e.,
$\overline{\varepsilon}(\vec{r},\omega)$ $\!\equiv$ $\!1$,
$\overline{\tens{H}}$ in Eq.~(\ref{eq28.12}) becomes
$\tens{H}_\mathrm{V}$ [recall Eqs.~(\ref{eq16})--(\ref{eq16.2})], 
leading to agreement with earlier results \cite{Power85,Power94}.

Bearing in mind the general relation (\ref{eq28.1.1}) between the CP
potential and many-atom vdW potentials, explicit expressions for
the two- and three-atom vdW potentials on vacuum background
can easily be obtained from formulas given in Sec.~\ref{sec1}
together with Eq.~(\ref{eq28}). {F}rom Eqs.~(\ref{eq17}),
(\ref{eq21}), and (\ref{eq24}) one can infer the well-known result
\cite{Casimir48} 
\begin{align}
\label{eq32}
&U_{AB}(\vec{r}_1,\vec{r}_2)
\nonumber\\&\quad
=-\frac{\hbar}{32\pi^3\varepsilon_0^2r^6}
 \int_0^{\infty} \mathrm{d} u\,g_2(ur/c) 
 \alpha_A(iu)\alpha_B(iu) 
\end{align}
($r$ $\!\equiv$ $\!|\vec{r}_1-\vec{r}_2|$), recall Eq.~(\ref{eq18}),
which reduces to  
\begin{equation}
\label{eq33}
U_{AB}(\vec{r}_1,\vec{r}_2)
=-\frac{23\hbar c\alpha_A(0)\alpha_B(0)}
 {64\pi^3\varepsilon_0^2r^7}
\end{equation}
in the retarded limit,
\begin{equation}
\label{eq33.1}
r_-\gg
 \frac{c}{\omega_-}\,,\quad
\end{equation}
with $r_-$ $\!\equiv$ $\!r$ and $\omega_-$ $\!\equiv$
$\!\mathrm{min}\{\omega_\mathrm{A}^-,\omega_\mathrm{B}^-\}$, and to 
\begin{equation}
\label{eq34}
U_{AB}(\vec{r}_1,\vec{r}_2)
=-\frac{3\hbar}
 {16\pi^3\varepsilon_0^2r^6}
\int_0^{\infty} \mathrm{d} u\, \alpha_A(iu)
 \alpha_B(iu)
\end{equation}
in the nonretarded limit,
\begin{equation}
\label{eq34.1}
r_+\ll
 \frac{c}{\omega_+}\,,\quad
\end{equation}
with $r_+$ $\!\equiv$ $\!r$ and $\omega_+$ $\!\equiv$
$\!\mathrm{max}\{\omega_\mathrm{A}^+,\omega_\mathrm{B}^+\}$.

Similarly, Eq.~(\ref{eq24.4}) implies that 
\begin{align}
\label{eq35}
&U_{ABC}(\vec{r}_1,\vec{r}_2,\vec{r}_3)
 =\frac{\hbar}{64\pi^4\varepsilon_0^3r_{12}^3r_{23}^3r_{31}^3}
 \int_0^{\infty} \mathrm{d} u 
 \nonumber\\
&\quad\times 
 \alpha_A(iu)
  \alpha_B(iu)\alpha_C(iu)
 g_3(u,\vec{r}_{12},\vec{r}_{23},\vec{r}_{31})
\end{align}
($\vec{r}_{ij}$ $\!\equiv$ $\!\vec{r}_i-\vec{r}_j$, $r_{ij}$
$\!\equiv$ $\!|\vec{r}_{ij}|$ for $i,j$ $\!=$ $\!1,2,3$), recall
Eq.~(\ref{eq24.5}), in agreement with the result found in
Refs.~\cite{Aub60,Power85,Power94}. Equations~(\ref{eq24.14}) and
(\ref{eq24.19}) show that Eq.~(\ref{eq35}) simplifies to
\begin{align}
\label{eq36}
&U_{ABC}(\vec{r}_1,\vec{r}_2,\vec{r}_3)
= \frac{\hbar c\alpha_A(0)\alpha_B(0)\alpha_C(0)}
 {16\pi^4\varepsilon_0^3
 r_{12}^3r_{23}^3r_{31}^3(r_{12}\!+\!r_{23}\!+\!r_{31})}
 \nonumber\\
&\times\!
 \bigl[f_1(r_{12},r_{23},r_{31})
 \!+\!f_2(r_{31},r_{12},r_{23})
 (\hat{\vec{r}}_{12}\!\cdot\!\hat{\vec{r}}_{23})^2\nonumber\\
&+\!f_2(r_{12},r_{23},r_{31})
 (\hat{\vec{r}}_{23}\!\cdot\!\hat{\vec{r}}_{31})^2
 \!\!+\!f_2(r_{23},r_{31},r_{12})
 (\hat{\vec{r}}_{31}\!\cdot\!\hat{\vec{r}}_{12})^2\nonumber\\
&+\!f_3(r_{12},r_{23},r_{31})
 (\hat{\vec{r}}_{12}\!\cdot\!\hat{\vec{r}}_{23})
 (\hat{\vec{r}}_{23}\!\cdot\!\hat{\vec{r}}_{31})
 (\hat{\vec{r}}_{31}\!\cdot\!\hat{\vec{r}}_{12})\bigr]
\end{align}
[$\hat{\vec{r}}_{ij}$ $\!\equiv$ $\vec{r}_{ij}/r_{ij}$, recall
Eq.~(\ref{eq24.10}) and Eqs.~(\ref{eq24.15})--(\ref{eq24.17})] in the
retarded limit [Eq.~(\ref{eq33.1}) with $r_-$ $\!\equiv$
$\!\min\{r_{12},$ $\!r_{23},$ $\!r_{31}\}$ and $\omega_-$
$\!\equiv$ $\!\min\{\omega_A^-,\omega_B^-,\omega_C^-\}$], and reduces
to the Axilrod-Teller potential \cite{Axilrod43}
\begin{align}
\label{eq37}
&U_{ABC}(\vec{r}_1,\vec{r}_2,\vec{r}_3)
\nonumber\\&\quad
=\frac{3\hbar\bigl[1-3
 (\hat{\vec{r}}_{12}\!\cdot\!\hat{\vec{r}}_{23})
 (\hat{\vec{r}}_{23}\!\cdot\!\hat{\vec{r}}_{31})
 (\hat{\vec{r}}_{31}\!\cdot\!\hat{\vec{r}}_{12})\bigr]}
 {64\pi^4\varepsilon_0^3r_{12}^3r_{23}^3r_{31}^3}
 \nonumber\\&\qquad
 \times\int_0^{\infty} \mathrm{d} u\, 
  \alpha_A(iu)\alpha_B(iu)\alpha_C(iu)
\end{align}
in the nonretarded limit [Eq.~(\ref{eq34.1}) with $r_+$ $\!\equiv$
$\!\max\{r_{12},$ $\!r_{23},$ $\!r_{31}\}$ and $\omega_+$
$\!\equiv$ $\!\max\{\omega_A^+,\omega_B^+,\omega_C^+\}$].


\section{Application to specific geometries}
\label{sec2}


\subsection{Dielectric ring}
\label{sec2.1}

\begin{figure}[!t!]
\noindent
\begin{center}
\includegraphics[width=\linewidth]{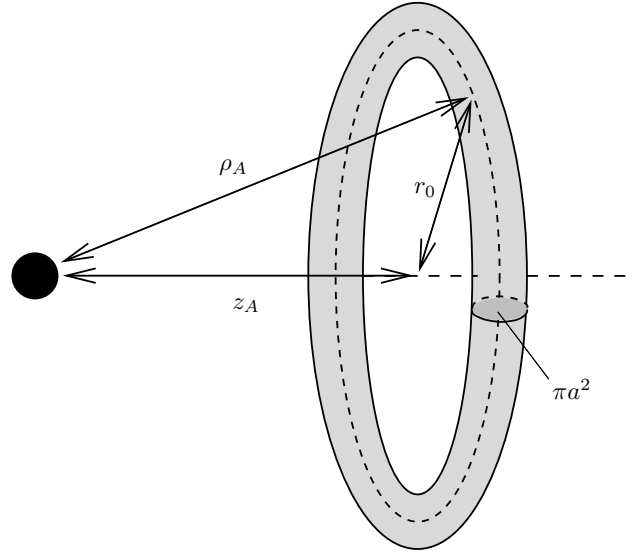}
\end{center}
\caption{
\label{fig1}
An atom near a dielectric ring (schematic picture).}
\end{figure}

Let us use the Born series given in Sec.~\ref{sec1} to calculate the 
(leading contributions to the) CP potential for some specific
geometries and begin with a ground-state atom placed on the
symmetry axis of a homogeneous dielectric ring of susceptibility
$\chi(\omega)$, having radius $r_0$, (circular) cross section $\pi
a^2$ (where $a$ $\!\ll$ $r_0$), and volume $V$ $\!=$ $\!2\pi^2 r_0
a^2$, the atom being separated from the center of the ring by a
distance $z_A$ (Fig.~\ref{fig1}).
From Fig.~\ref{fig1} we see that $|\mathbf{r}_A-\mathbf{s}|$
$\!\simeq$ $\!\sqrt{z_A^2+r_0^2}$ $\!\equiv$ $\!\rho_A$ for $a$
$\!\ll$ $\!r_0$, so an evaluation of the (trivial) volume integral in
Eq.~(\ref{eq17}) results in the first-order CP potential
\begin{align}
\label{eq25}
&\Delta_1U_A(\rho_A)
\nonumber\\&\quad
= -\frac{\hbar V}{32\pi^3\varepsilon_0\rho_A^6}
 \int_0^{\infty} \mathrm{d} u\,\alpha_A(iu)\chi(iu)g_2(u\rho_A/c),
\end{align}
which is attractive, as expected. In the retarded limit
[Eq.~(\ref{eq19}) with $r_-$ $\!=$ $\!\rho_A$] Eq.~(\ref{eq25})
reduces to
\begin{equation}
\label{eq26}
\Delta_1U_A(\rho_A)
=-\frac{23\hbar cV\alpha_A(0)\chi(0)}
 {64\pi^3\varepsilon_0\rho_A^7}
\end{equation}
[cf. Eq.~(\ref{eq21})], while in the nonretarded limit
[Eq.~(\ref{eq22}) with $r_+$ $\!=$ $\!\rho_A$] one easily finds
\begin{equation}
\label{eq27}
\Delta_1U_A(\rho_A)
=-\frac{3\hbar V}
 {16\pi^3\varepsilon_0\rho_A^6}
 \int_0^{\infty} \mathrm{d} u\, \alpha_A(iu)
 \chi(iu)
\end{equation}
[cf. Eq.~(\ref{eq24})]. In both limiting cases the CP potential thus
reduces to simple asymptotic power laws in $\rho_A$, where as usual
the leading (inverse) power is increased by one when going from the
nonretarded to the retarded limit.

The first (single point) second-order correction term
$\Delta_2^1U_A(\rho_A)$ can simply be obtained from 
Eqs.~(\ref{eq25})--(\ref{eq27}) by means of the replacement
(\ref{eq24.3}), while the calculation of the second (two-point) term
$\Delta_2^2U_A(\rho_A)$ is a lot more difficult due to the factor
$|\mathbf{s}_1$ $\!-$ $\!\mathbf{s}_2|$. We find (see
App.~\ref{AppB})
\begin{equation}
\label{eq27.10}
\Delta_2^2U_A(\rho_A)
=\frac{(0.05\pm 0.02)\hbar cV\alpha_A(0)\chi^2(0)}
 {\pi^3\varepsilon_0\rho_A^7}
\end{equation}
in the retarded limit and
\begin{equation}
\label{eq27.11}
\Delta_2^2U_A(\rho_A)
=\frac{(0.08\pm 0.03)\hbar V}
 {\pi^3\varepsilon_0\rho_A^6}
 \int_0^{\infty} \mathrm{d} u\, \alpha_A(iu)
 \chi^2(iu)
\end{equation}
in the nonretarded limit. Recalling Eqs.~(\ref{eq24.1}) and
(\ref{eq24.3}), Eqs.~(\ref{eq26}) and (\ref{eq27.10}) imply that 
up to quadratic order in $\chi$ we have
\begin{equation}
\label{eq27.12}
 U_A(\rho_A) 
 =-\frac{23\hbar cV\alpha_A(0)\chi(0)}
 {64\pi^3\varepsilon_0\rho_A^7}
 \Big[1-(0.47\pm 0.05)\chi(0)\Big]
\end{equation}
in the retarded limit, while Eqs.~(\ref{eq27}) and (\ref{eq27.11})
show that in the nonretarded limit
\begin{align}
\label{eq27.13}
&U_A(\rho_A)
=-\frac{3\hbar V}
 {16\pi^3\varepsilon_0\rho_A^6}
 \nonumber\\
&\ \times
 \int_0^{\infty} \mathrm{d} u\, \alpha_A(iu)
 \chi(iu) \Big[1-(0.77\pm 0.17)\chi(iu)\Big].
\end{align}
The uncertainty in the magnitude of the contribution quadratic in
$\chi$ is due to the approximations made when calculating 
$\Delta^2_2U(\rho_A)$ (cf. App.~\ref{AppB}). However, irrespective of
these approximations, Eqs.~(\ref{eq27.12}) and (\ref{eq27.13}) show
that the leading non-additive correction $\Delta^2_2U_A(\rho_A)$ to
the linear result $\Delta_1U_A(\rho_A)$ does not change the powers in
the asymptotic retarded and nonretarded distance laws ($\rho_A^{-7}$
and $\rho_A^{-6}$, respectively), but merely modifies the constants of
proportionality. A similar result has been found when studying a
dielectric half space \cite{Buhmann05b}.


\subsection{Many-body decomposition}
\label{sec2.2}

The explicit evaluation of multiple spatial integrals [being the main
difficulty when evaluating $\Delta^2_2U(\vec{r}_A)$] can in fact be
avoided in many cases by an appropriate decomposition of the body of
interest, as shall be demonstrated in the following. To that end, let
us decompose the body described by $\chi(\vec{r},\omega)$ [recall 
Eq.~(\ref{eq7})] into smaller bodies numbered by $n$, so that
\begin{equation}
\label{eq10}
\chi(\vec{r},\omega)
=\sum_n\chi_n(\vec{r},\omega)1_{V_n}(\vec{r}),
\end{equation}
where
\begin{equation}
\label{eq11}
1_{V_n}(\vec{r})
=\left\{\begin{array}{l}1\mbox{ for }\vec{r}\in V_n,\\[.5ex]
0 \mbox{ for }\vec{r}\notin V_n. \end{array}\right.
\end{equation}
Substituting Eqs.~(\ref{eq10}) and (\ref{eq11}) into Eq.~(\ref{eq9}),
and slight\-ly rearranging the terms, we obtain
\begin{equation}
\label{eq11.1}
\Delta_kU_A(\vec{r}_A)
=\sum_{l=1}^k\Delta_k^lU_A(\vec{r}_A),
\end{equation}
where
\begin{equation}
\label{eq11.1.1}
\Delta_k^lU_A(\vec{r}_A)=
 \sum_{n_1<\ldots <n_l}
 \Delta_k^lU_A^{n_1\ldots n_l}(\vec{r}_A)
\end{equation}
with
\begin{equation}
\label{eq11.2}
\Delta_k^lU_A^{n_1\ldots n_l}(\vec{r}_A)
=\sum_{(m_1,\ldots,m_k)\in\mathcal{I}^k_{n_1\ldots n_l}}
 W_A^{m_1\ldots m_k}(\vec{r}_A)
\end{equation}
is the sum of all $l$-body contributions of order $k$ in $\chi$.
In Eq.~(\ref{eq11.2}),
\begin{align}
\label{eq11.4}
&W_A^{m_1\ldots m_k}(\vec{r}_A)
 =\frac{(-1)^k\hbar\mu_0}{2\pi c^{2k}}\nonumber\\
&\times\int_0^{\infty} \!\mathrm{d} u \,u^{2k+2}\alpha_A(iu)
 \Biggl[\prod_{j=1}^k \int_{V_{m_j}}\!\!\mathrm{d}^3s_j\,
 \chi_{m_j}(\vec{s}_j,iu)\Biggr]\nonumber\\
&\times \mathrm{Tr}\bigl[
 \overline{\tens{G}}(\vec{r}_A,\vec{s}_1,iu)
 \cdot\overline{\tens{G}}(\vec{s}_1,\vec{s}_2,iu)\cdots
\overline{\tens{G}}(\vec{s}_k,\vec{r}_A,iu)\bigr],
\end{align}
and the notation
\begin{align}
\label{eq11.3}
&\mathcal{I}^k_{n_1\ldots n_l}\nonumber\\
&=\Bigl\{(m_1,\ldots,m_k)\in\{n_1,\ldots,n_l\}^k
|\;\forall i\;\exists j:m_j=n_i\Bigr\}
\end{align}
is used.

In particular, to linear order in $\chi$ we have
\begin{align}
\label{eq12}
\Delta_1U_A(\vec{r}_A)&\,=\Delta_1^1U_A(\vec{r}_A)
=\sum_n\Delta_1^1U_A^n(\vec{r}_A)\nonumber\\
&\,=\sum_nW_A^n(\vec{r}_A),
\end{align}
so the CP potential is additive in this order. For the term quadratic
in $\chi$, Eqs.~(\ref{eq11.1})--(\ref{eq11.2}) together with
Eq.~(\ref{eq11.3}) reduce to
\begin{align}
\label{eq13}
&\Delta_2U_A(\vec{r}_A)
=\Delta_2^1U_A(\vec{r}_A)+\Delta_2^2U_A(\vec{r}_A)
 \nonumber\\
&\quad=\sum_n\Delta_2^1U_A^n(\vec{r}_A)
 +\sum_{m<n}\Delta_2^2U_A^{mn}(\vec{r}_A)
 \nonumber\\
&\quad=\sum_nW_A^{nn}(\vec{r}_A)
 +\sum_{m<n}\bigl[W_A^{mn}(\vec{r}_A)
 +W_A^{nm}(\vec{r}_A)\bigr].
\end{align}
Obviously, the second term on the r.h.s. of Eq.~(\ref{eq13}) is the
(overall) two-body contribution to the CP potential up to quadratic
order in $\chi$. It can be regarded as being the leading correction to
the additivity of the potential. Clearly, the expansion can in
principle be extended to arbitrarily high orders in $\chi$, whereby
$k$-body interactions first appear at $k$th order in $\chi$. In
particular, Eqs.~(\ref{eq12}) and (\ref{eq13}) generalize the result
that up to linear order in $\chi$ the CP potential of an atom near a
homogeneous semi-infinite dielectric half space can be written as an
(infinite) sum of thin-layer potentials \cite{Buhmann05}, whereas the
contribution quadratic in $\chi$ also contains two-layer terms leading
to a breakdown of additivity \cite{Buhmann05b}.

To illustrate the application of Eqs.~(\ref{eq11.1})--(\ref{eq11.3}) 
to the calculation of the CP potential of complex bodies via
decomposition into simpler bodies, let us consider an atom at position
$-z_A$ ($z_A$ $\!>$ $0$) near a semi-infinite half space filled with a
stratified dielectric medium, i.e.
\begin{equation}
\label{eq13.1}
\varepsilon(\vec{r},\omega)=1+\chi(\omega)p(z),
\end{equation}
where $p(z)$ is some profile function [$p(z)$ $\!\ge$ $\!0$ for $z$
$\!\ge$ $\!0$, $p(z)$ $\!=$ $\!0$ for $z$ $\!<$ $\!0$], which may be
normalized such that $\max p(z)$ $\!=$ $\!1$. We decompose the half
space into a set of thin slices of equal thickness $d$ such that 
\begin{equation}
\label{eq13.2}
d\max\{p'(z)|z>0\}\ll 1.
\end{equation}
{F}rom Eq.~(\ref{eq12}) it follows that $\Delta_1^1U_A(z_A)$ is given by
the sum over the slices, which for $d$ $\!\ll$ $z_A$ contribute
\begin{align}
\label{eq13.3}
&\Delta_1^1U_A^n(z_A)=-\frac{\hbar\mu_0d}{4\pi^2}
 \int_0^{\infty}\!\!\mathrm{d}u\,u^2 \alpha_A(iu)
 \int_0^\infty\!\!\mathrm{d}q\,q
 \nonumber\\
&\times e^{-2b(z_\mathrm{A}+nd)}
\biggl[\biggl(\frac{bc}{u}\biggr)^{\!\!2}\hspace{-0.5ex}-1
+\frac{1}{2}\biggl(\frac{u}{bc}\biggr)^{\!\!2}\biggr]\chi(iu)p(nd),
\end{align}
where
\begin{equation}
\label{eq13.3.1}
 b = \sqrt{\frac{u^2}{c^2}+q^2}
\end{equation}
(cf. Eq.~(71) in Ref.~\cite{Buhmann05b}), so after turning the sum
into an integral one obtains
\begin{align}
\label{eq13.4}
&\Delta_1^1U_A(z_A)=-\frac{\hbar\mu_0}{4\pi^2}
 \int_0^{\infty}\!\!\mathrm{d}u\,u^2 \alpha_A(iu)\chi(iu)
 \nonumber\\
&\quad\times 
 \int_0^\infty\!\!\mathrm{d}q\,q
 e^{-2bz_\mathrm{A}}P(2b)
\biggl[\biggl(\frac{bc}{u}\biggr)^{\!\!2}\hspace{-0.5ex}-1
+\frac{1}{2}\biggl(\frac{u}{bc}\biggr)^{\!\!2}\biggr],
\end{align}
where
\begin{equation}
\label{eq13.5}
P(x)=\int_0^\infty\mathrm{d}z\,e^{-xz}p(z)
\end{equation}
is the Laplace transform of the profile function $p(z)$. In a similar
way, the results
\begin{align}
\label{eq13.6}
&\Delta_2^1U_A^n(z_A)=\frac{\hbar\mu_0d}{4\pi^2}
 \!\int_0^{\infty}\!\!\mathrm{d}u\,u^2 \alpha_A(iu)
 \!\int_0^\infty\!\!\mathrm{d}q\,q\nonumber\\
&\times e^{-2b(z_\mathrm{A}+nd)}
\biggl[\frac{1}{2}\biggl(\frac{bc}{u}\biggr)^{\!\!2}
 \hspace{-0.5ex}-\frac{3}{4}
 +\frac{1}{4}\biggl(\frac{u}{bc}\biggr)^{\!\!2}
 \biggr]\chi^2(iu)p(nd)
\end{align}
(cf. Eq.~(74) in Ref.~\cite{Buhmann05b}) and
\begin{align}
\label{eq13.7}
&\Delta_2^2U_A^{mn}(z_A)=\frac{\hbar\mu_0d^2}{2\pi^2}
 \!\int_0^{\infty}\!\!\mathrm{d}u\,u^2 \alpha_A(iu)
 \!\int_0^\infty\!\!\mathrm{d}q\,qb\nonumber\\
&\quad\times e^{-2b(z_\mathrm{A}+nd)}
\biggl[\frac{1}{2}
 -\frac{1}{2}\biggl(\frac{u}{bc}\biggr)^{\!\!2}
 +\frac{1}{4}\biggl(\frac{u}{bc}\biggr)^{\!\!4}
 \biggr]\nonumber\\
&\qquad\times\chi(iu)p(md)\chi(iu)p(nd)
\end{align}
(cf. Eq.~(75) in Ref.~\cite{Buhmann05b}) can be derived, leading to
\begin{align}
\label{eq13.8}
&\Delta_2^1U_A(z_A)=\frac{\hbar\mu_0}{4\pi^2}
 \int_0^{\infty}\!\mathrm{d}u\,u^2 \alpha_A(iu)\chi^2(iu)
 \nonumber\\
&\times\int_0^\infty\!\mathrm{d}q\,q
 e^{-2bz_\mathrm{A}}P(2b)
 \biggl[\frac{1}{2}\biggl(\frac{bc}{u}\biggr)^{\!\!2}
 -\frac{3}{4}
 +\frac{1}{4}\biggl(\frac{u}{bc}\biggr)^{\!\!2}
 \biggr]
\end{align}
and
\begin{align}
\label{eq13.9}
&\Delta_2^2U_A(z_A)=\frac{\hbar\mu_0}{2\pi^2}
 \int_0^{\infty}\!\mathrm{d}u\,u^2 \alpha_A(iu)\chi^2(iu)
 \nonumber\\
&\quad\times\int_0^\infty\!\mathrm{d}q\,qb
e^{-2bz_\mathrm{A}}
\biggl[\frac{1}{2}
 -\frac{1}{2}\biggl(\frac{u}{bc}\biggr)^{\!\!2}
 +\frac{1}{4}\biggl(\frac{u}{bc}\biggr)^{\!\!4}
 \biggr]\nonumber\\
&\qquad\quad\times
 \int_0^\infty\mathrm{d}z\,p(z)
 \int_z^\infty\mathrm{d}z'\,e^{-2bz'}p(z'),
\end{align}
respectively. Hence the CP potential of the inhomogeneous half space
up to quadratic order of $\chi$ has been calculated from the known CP
potentials of one and two thin plates of constant permittivities.
Needless to say that the method can be carried out to higher orders of
$\chi$ and can also be applied to other than planar systems.

To give an example, we consider a dielectric medium whose permittivity
oscillates in the $z$ direction, 
\begin{equation}
\label{eq13.10}
\varepsilon(\vec{r},\omega)=1+\chi(\omega)\cos^2(k_zz)\Theta(z)
\end{equation}
[$\Theta(z)$, unit step function]. Using Eqs.~(\ref{eq13.4}),
(\ref{eq13.8}), and (\ref{eq13.9}), we find that up to quadratic order
in $\chi$ the CP potential takes the asymptotic form (see
App.~\ref{AppA})
\begin{align}
\label{eq13.15}
&U_A(z_A)
= -\frac{1}{z_A^4}\Bigl\{\Delta_1C_4F_3(k_zz_A)\nonumber\\
&\qquad
 +\Delta_2C_4\Bigl[{\textstyle\frac{126}{169}}F_3(k_zz_A)
 +{\textstyle\frac{43}{169}}H_3(k_zz_A)\Bigr]\Bigr\}
\end{align}
in the retarded limit [Eq.~(\ref{eq19}) with $r_-$ $\!=$ $z_A$] and
\begin{equation}
\label{eq13.16}
U_A(z_A) = -\frac{(\Delta_1C_3+\Delta_2C_3)F_2(k_zz_A)}{z_A^3}
\end{equation}
in the nonretarded limit [Eq.~(\ref{eq22}) with $r_+$ $\!=$ $z_A$].
Here,
\begin{align}
\label{A3}
\Delta_1C_4
=&\,\frac{23\hbar c\alpha_A(0)\chi(0)}{640\pi^2\varepsilon_0}\,,\\
\label{A4}
\Delta_2C_4
=&\,-\frac{169\hbar c\alpha_A(0)\chi^2(0)}{8960\pi^2\varepsilon_0}\,,
\end{align} 
\begin{align}
\label{A5}
\Delta_1C_3
=&\,\frac{\hbar}{32\pi^2\varepsilon_0}
 \int_0^\infty\mathrm{d}u\,\alpha_A(iu)\chi(iu)\,,\\
\label{A6}
\Delta_2C_3
=&\,-\frac{\hbar}{64\pi^2\varepsilon_0}
 \int_0^\infty\mathrm{d}u\,\alpha_A(iu)\chi^2(iu)
\end{align}
are just the linear and and quadratic expansions of the well-known
coefficients for the homogeneous half space in the retarded and
nonretarded limits [where $U_A(z_A)$ $\!=$ $\!-C_4/z_A^4$, and
$U(z_A)$ $\!=$ $\!-C_3/z_A^3$, respectively], and the structure
functions
\begin{align}
\label{eq13.17}
&F_j(x) = \frac{2^j}{j!}\int_0^\infty\mathrm{d}t\,t^je^{-2t}
 \,\frac{2t^2+x^2}{t^2+x^2}\,,\\
\label{eq13.18}
&H_j(x)= \frac{2^j}{j!}\int_0^\infty\!\mathrm{d}t\,t^je^{-2t}
 \,\frac{2t^6\!+\!8x^2t^4\!+\!5x^4t^2\!+\!2x^6}
   {(t^2\!+\!x^2)^2(t^2\!+\!4x^2)}
\end{align}
are normalized such that
\begin{align}
\label{eq13.19}
&F_j(x)\to\biggl\{\begin{array}{lcl}1&\mathrm{for}&x\to 0,\\
 {\textstyle\frac{1}{2}}&\mathrm{for}&x\to\infty,\end{array}
 \\
\label{eq13.20}
&H_j(x)\to\biggl\{\begin{array}{lcl}1&\mathrm{for}&x\to 0,\\
 {\textstyle\frac{1}{4}}&\mathrm{for}&x\to\infty.
\end{array}
\end{align}

\begin{figure}[!t!]
\noindent
\begin{center}
\includegraphics[width=\linewidth]{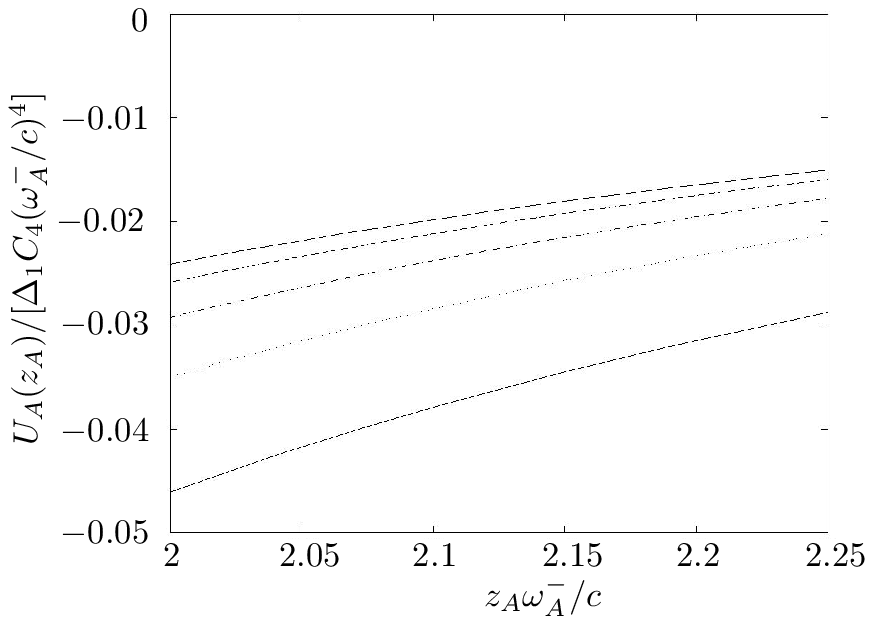}
\end{center}
\caption{
\label{fig2}
The CP potential of a ground-state atom in front of a half space with
oscillating susceptibility in the retarded limit is displayed as a
function of the distance $z_A$, where $k_zc/\omega_A^-$ $\!\to$
$\!\infty$ (upper solid line), $k_zc/\omega_A^-$ $\!=$ $\!4$ (dashed
line), $k_zc/\omega_A^-$ $\!=$ $\!2$ (dash-dotted line),
$k_zc/\omega_A^-$ $\!=$ $\!1$ (dotted line), $k_zc/\omega_A^-$ $\!\to$
$\!0$ (lower solid line), $\chi(0)$ $\!=$ $\!\frac{1}{2}$.}
\end{figure}

\begin{figure}[!t!]
\noindent
\begin{center}
\includegraphics[width=\linewidth]{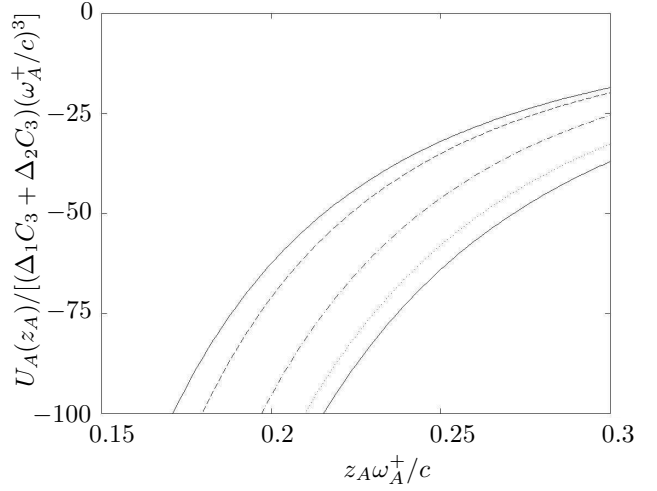}
\end{center}
\caption{
\label{fig3}
The CP potential of a ground-state atom in front of a half space with
oscillating susceptibility in the non-retarded limit is displayed as a
function of the distance $z_A$, where $k_zc/\omega_A^+$ $\!\to$
$\!\infty$ (upper solid line), $k_zc/\omega_A^+$ $\!=$ $\!20$ (dashed
line), $k_zc/\omega_A^+$ $\!=$ $\!6$ (dash-dotted line),
$k_zc/\omega_A^+$ $\!=$ $\!2$ (dotted line), $k_zc/\omega_A^-$ $\!\to$
$\!0$ (lower solid line).}
\end{figure}

Equations (\ref{eq13.15}) and (\ref{eq13.16}), respectively, are 
illustrated in Figs.~\ref{fig2} and \ref{fig3}.
It is seen that the potential curves for different values of $k_z$
lie between the two solid curves that correspond to the limiting
cases $k_z$ $\!\to$ $\!\infty$ (upper curves) and $k_z$ $\!\to$ $\!0$
(lower curves, which represent the potential observed in the case of
the respective homogeneous half space). {F}rom Eqs.~(\ref{eq13.15}),
(\ref{eq13.16}), and (\ref{eq13.19}) it follows that the upper-curve
potential values obtained in linear order of $\chi$ are $1/2$ times
the lower-curve ones, which reflects the fact that for $k_z$ $\!\to$
$\!\infty$ the potential in linear order of $\chi$ is simply
determined by the average permittivity $\varepsilon(\vec{r},\omega)$
$\!\simeq$ $\!1+\frac{1}{2}\chi(\omega)\Theta(z)$, cf.
Eq.~(\ref{eq13.10}). The factor found for the quadratic-order term is
equal to $1/2$ in the non-retarded limit [cf. Eqs.~(\ref{eq13.16}) and
(\ref{eq13.19})], but equal to $295/676$ in the retarded limit [cf.
Eqs.~(\ref{eq13.15}), (\ref{eq13.19}), and (\ref{eq13.20})], owing to
the influence of the two-plate term $\Delta_2^2U_A(z_A)$. Note that
the curves for the intermediate values of $k_z$ approach the upper
limiting curve for large values of $z_A$ and the lower limiting curve
for small values of $z_A$, the potentials thus being (near $z_A$
$\!\simeq$ $k_z^{-1}$) somewhat steeper than $z_A^{-4}$ and
$z_A^{-3}$---the power laws observed in the case of a homogeneous half
space. By controlling $k_z$, one can therefore control the shape of
the potentials.


\section{Summary}
\label{sec4}

Within leading-order perturbation theory, the CP potential of a
ground-state atom near dielectric bodies can be expressed in terms of
the atomic polarizability and the scattering Green tensor of the
body-assisted electromagnetic field, where the bodies are
characterized by a spatially varying dielectric susceptibility that is
a complex function of frequency. Starting from this very general
formula, we have performed a Born expansion of the Green tensor to
obtain an expansion of the CP potential in powers of the electric
susceptibility. The expansion shows that only in linear order the CP
force is a sum of attractive central forces, while higher-order terms
are unavoidably connected to multiple-point correlations in the
dielectric matter, leading to a breakdown of additivity. 

Using the Born series, we have shown that when the dielectric bodies
can be described by a susceptibility of Clausius-Mosotti type, i.e.,
when the basic constituents can be regarded as atom-like, then the CP
potential is the (infinite) sum of all microscopic many-atom vdW
potentials between the atom under consideration and the atoms forming
the bodies. As a by-product, a general formula for the many-atom vdW
potential of arbitrary order and on an arbitrary background of
dielectric bodies has been found, which generalizes previous results
found for atoms in vacuum.  

Apart from being useful for making contact with microscopic
descriptions of the CP force, the Born series can also be used for
practical calculations, particularly when the Green tensor is not
available in closed form. We have employed two strategies. (i) By
direct evaluation of multiple spatial integrals, we have determined
the attractive CP potential of a weakly dielectric ring, finding
asymptotic $1/\rho_A^7$ and $1/\rho_A^6$ power laws in the retarded
and the nonretarded limits, respectively. (ii) By reduction to simpler
bodies with known CP potentials, we have derived expressions for the
CP potential of an atom placed in front of an inhomogeneous stratified
half space, with special emphasis on an oscillating susceptibility. In
this case the potential exhibits---for distances comparable to the
oscillation period---a somewhat stronger power law than in the case of
a homogeneous half space.

\begin{acknowledgement}
This work was supported by the Deu\-tsche Forschungsgemeinschaft.
S.Y.B. would like to thank Gabriel Barton, Christian Raabe, and Ho
Trung Dung for discussions.
\end{acknowledgement}


\appendix


\section{Derivation of the expansion in terms of many-atom vdW
potentials [Eq.~(\ref{eq28.1.1})]}
\label{AppC}

As a preparation, we derive the symmetrization (\ref{eq28.5}). The
completely symmetrized form of a many-atom potential is given by
\begin{equation}
\label{eq28.2}
\mathcal{S}f(\vec{r}_1,\ldots,\vec{r}_j)
 \equiv\frac{1}{(2-\delta_{2j})j}\sum_{\pi\in P(j)}
 f(\vec{r}_{\pi(1)},\ldots,\vec{r}_{\pi(j)}),
\end{equation}
where $P(j)$ denotes the permutation group of the numbers $1,\ldots,j$
and $1/(2-\delta_{2j})j$ is a normalization factor. As a trivial
consequence of the cyclic property of the trace as well as the
symmetry property of the Green tensor \cite{Knoll01}
\begin{equation}
\label{eq28.3}
\tens{G}(\vec{r},\vec{r}',\omega)
=\tens{G}^\top(\vec{r}',\vec{r},\omega)
\end{equation}
together with Eq.~(\ref{eq28.1}) one easily finds that
\begin{align}
\label{eq28.4}
&\mathrm{Tr}\bigl[
 \overline{\tens{H}}(\vec{r}_1,\vec{r}_2,\omega)\cdots
 \overline{\tens{H}}(\vec{r}_j,\vec{r}_1,\omega)\bigr]\nonumber\\
&\quad =\mathrm{Tr}\bigl[
 \overline{\tens{H}}
 (\vec{r}_{\pi(1)},\vec{r}_{\pi(2)},\omega)\cdots
 \overline{\tens{H}}
 (\vec{r}_{\pi(j)},\vec{r}_{\pi(1)},\omega)\bigr]
\end{align}
if $\pi$ is either a cyclic permutation [e.g., $\pi(1)$ $\!=$ $\!2$,
$\pi(2)$ $\!=$ $\!3$, $\ldots$ , $\pi(j)$ $\!=$ $\!1$] or the reverse
of a cyclic permutation [e.g., $\pi(1)$ $\!=$ $\!j$,
$\pi(2)$ $\!=$ $\!j-1$, $\ldots$ , $\pi(j)$ $\!=$ $\!1$]. With
$f(\vec{r}_1,\ldots,\vec{r}_j)$ being given by the
l.h.s. of Eq.~(\ref{eq28.4}), the sum on the r.h.s. of
Eq.~(\ref{eq28.2}) contains classes of $(2-\delta_{2j})j$ terms that
give the same result (note that for $j$ $\!=$ $\!2$ the cyclic
permutation and its reverse coincide, so we have only $j$ instead
of $2j$ terms in the class). By forming a set $\overline{P}(j)$
$\!\varsubsetneq$ $\!P(j)$ containing exactly one representative
of each class (where obviously $\overline{P}(j)$ has $j!/$
$\![(2\!-\!\delta_{2j})j]$ members), the sum can thus be simplified,
leading to Eq.~(\ref{eq28.5}).

With this preparation at hand, we may derive Eq. (\ref{eq28.1.1})
[together with Eqs.~(\ref{eq28.11}) and (\ref{eq28.5})] by following
these steps: We substitute Eqs. (\ref{eq28}) and (\ref{eq28.1}) into
Eq.~(\ref{eq9}), multiply out and perform all spatial integrals over
delta functions, where only terms of the form
$\delta(\vec{s}_j-\vec{s}_{j+1})$ contribute [terms of the
form $\delta(\vec{r}_A-\vec{s}_j)$ giving zero integrals, because of
$n(\vec{r}_A)$ $\!=$ $\!0$, cf. the remark below Eq.~(\ref{eq28})].
After renaming the remaining integration variables according to 
\begin{align}
\label{C1}
&\Biggl[\prod_{j=1}^l\int\mathrm{d}^3s_j\,n(\vec{s}_j)\Biggr]
 f(\vec{s}_1,\ldots,\vec{s}_l)\nonumber\\
&=\frac{1}{l!}\sum_{\pi\in P(l)}
 \Biggl[\prod_{j=1}^l\int\mathrm{d}^3s_j\,n(\vec{s}_j)\Biggr]
 f(\vec{s}_{\pi(1)},\ldots,\vec{s}_{\pi(l)}),
\end{align}
the result may be written in the form
\begin{equation}
\label{eq28.6}
\Delta_kU_A(\vec{r}_A)
=\sum_{l=1}^k\Delta_k^lU_A(\vec{r}_A),
\end{equation}
with
\begin{align}
\label{eq28.7}
\Delta_k^lU_A(\vec{r}_A)
=&\,\frac{1}{l!}\int_0^\infty\mathrm{d}u
 \Biggl[\prod_{j=1}^l\int
 \frac{\mathrm{d}^3s_j\,n(\vec{s}_j)}
 {1\!-\!\frac{1}{3}\varepsilon_0^{-1}n(\vec{s}_j)\alpha_B(\omega)}
 \Biggr]\nonumber\\
&\times\!\!\!
\sum_{\begin{array}{c}\scriptstyle\eta_1\ge 0,\ldots,\eta_l\ge 0\\
 \scriptstyle\eta_1+\ldots+\eta_l=k-l\end{array}}
 \hspace{-3ex}q^{\eta_1}(\vec{s}_1,iu)\cdots
 q^{\eta_l}(\vec{s}_l,iu)\nonumber\\
&\times U_{AB\ldots B}(\vec{r}_A,\vec{s}_1,\ldots,\vec{s}_l,iu),
\end{align}
where each power of the factor
\begin{equation}
\label{C2}
q(\vec{r},\omega)=
 \frac{-\frac{1}{3}\varepsilon_0^{-1}n(\vec{r})\alpha_B(\omega)}
 {1-\frac{1}{3}\varepsilon_0^{-1}n(\vec{r})\alpha_B(\omega)}
\end{equation}
is due to the integration of one term containing
$\delta(\vec{s}_j-\vec{s}_{j+1})$, and 
\begin{align}
\label{C3}
&\int_0^\infty\mathrm{d}u\,
 U_{AB\ldots B}(\vec{r}_1,\ldots,\vec{r}_{l+1},iu)
 \nonumber\\
&=U_{AB\ldots B}(\vec{r}_1,\ldots,\vec{r}_{l+1}),
\end{align}
recall Eq.~(\ref{eq28.11}). Summing Eq.~(\ref{eq28.6}) over $k$, and
rearranging the double sum, we find
\begin{equation}
\label{C4}
\sum_{k=1}^\infty\Delta_kU_A(\vec{r}_A)
=\sum_{k=1}^\infty\sum_{l=1}^k\Delta_k^lU_A(\vec{r}_A)
=\sum_{l=1}^\infty\Delta^lU_A(\vec{r}_A),
\end{equation}
where
\begin{align}
\label{C5}
&\Delta^lU_A(\vec{r}_A)
=\frac{1}{l!}\int_0^\infty\!\!\mathrm{d}u\,
 \Biggl[\prod_{j=1}^l\int\mathrm{d}^3s_j\,\nonumber\\
&\quad\times\frac{n(\vec{s}_j)}
 {1-\frac{1}{3}\varepsilon_0^{-1}n(\vec{s}_j)\alpha_B(iu)}
 \sum_{\eta_j=0}^\infty q^{\eta_j}(\vec{s}_j,iu)\Biggr]
\nonumber\\
&\quad\times
 U_{AB\ldots B}(\vec{r}_A,\vec{s}_1,\ldots,\vec{s}_l,iu).
\end{align}
After performing the geometric sums
\begin{align}
\label{C6}
\sum_{j=0}^\infty q^j(\vec{r},\omega)
 =1-{\textstyle\frac{1}{3}}\varepsilon_0^{-1}
 n(\vec{r})\alpha_B(\omega),
\end{align}
cf. Eq.~(\ref{C2}), the denominators in Eq.~(\ref{C5}) cancel, so by
recalling Eqs.~(\ref{eq9.1}), (\ref{C3}) and (\ref{C4}), we arrive at
Eq.~(\ref{eq28.1.1}) together with Eqs.~(\ref{eq28.11}) and
(\ref{eq28.5}).


\section{Calculation of the two-point correlation term for
the dielectric ring [Eqs.~(\ref{eq27.10}) and (\ref{eq27.11})]}
\label{AppB}

An approximation to the two-point correlation term in the retarded
limit as given by Eq.~(\ref{eq24.14})--(\ref{eq24.17}) [together with
Eqs.~(\ref{eq24.6})--(\ref{eq24.8}) and Eq.~(\ref{eq24.10})] in the
case of the dielectric ring can be obtained by replacing
the variable $\vec{s}_1$ by its average across the cross section of
the ring ($|\mathbf{r}_A-\mathbf{s}_1|$ $\!\simeq$ $\!\rho_A$ for $a$
$\!\ll$ $\!r_0$), evaluating the $\vec{s}_1$-integral, and separating
the $\vec{s}_2$-integral into two parts,
\begin{align}
\label{B1}
&\Delta_2^2U_A(\vec{r}_A)
=\frac{\hbar cV\alpha_A(0)\chi^2(0)}{32\pi^4\varepsilon_0}
\nonumber\\
&\quad\times\Biggl\{
 \int_{-\lambda a}^{\lambda a}\mathrm{d}z\int_0^a\mathrm{d}\rho\,\rho
 \int_0^{2\pi}\mathrm{d}\phi
 +\pi a^2\int_{\lambda a/r_0}^{2\pi-\lambda a/r_0}r_0\mathrm{d}\theta
 \Biggr\}\nonumber\\
&\quad\times
 \frac{1}
 {\alpha^3\beta^3\gamma^3(\alpha\!+\!\beta\!+\!\gamma)}
 \bigl[f_1(\alpha,\beta,\gamma)
 +f_2(\gamma,\alpha,\beta)
 (\hat{\bm{\alpha}}\!\cdot\!\hat{\bm{\beta}})^2\nonumber\\
&\quad+f_2(\alpha,\beta,\gamma)
 (\hat{\bm{\beta}}\!\cdot\!\hat{\bm{\gamma}})^2
 +f_2(\beta,\gamma,\alpha)
 (\hat{\bm{\gamma}}\!\cdot\!\hat{\bm{\alpha}})^2\nonumber\\
&\quad+f_3(\alpha,\beta,\gamma)
 (\hat{\bm{\alpha}}\!\cdot\!\hat{\bm{\beta}})
 (\hat{\bm{\beta}}\!\cdot\!\hat{\bm{\gamma}})
 (\hat{\bm{\gamma}}\!\cdot\!\hat{\bm{\alpha}})\bigr]\nonumber\\
&\equiv\Delta_2^{2,\mathrm{c}}U_A(\vec{r}_A)+
 \Delta_2^{2,\mathrm{r}}U_A(\vec{r}_A)
\end{align}
where the integral in $\Delta_2^{2,\mathrm{c}}U_A(\vec{r}_A)$ extends
over an approximately cylindrical volume of cross section $\pi a^2$
and length $2\lambda a$, and that
in $\Delta_2^{2,\mathrm{r}}U_A(\vec{r}_A)$ extends over the volume of
the remaining open ring.

For the integral in $\Delta_2^{2,\mathrm{c}}U_A(\vec{r}_A)$, we may
approximate
\begin{align}
\label{B2}
&\qquad\alpha=\gamma\simeq\rho_A,\quad
\beta\simeq \sqrt{z^2+\rho^2},\nonumber\\
&\hat{\bm{\alpha}}\!\cdot\!\hat{\bm{\beta}}
 =-\hat{\bm{\beta}}\!\cdot\!\hat{\bm{\gamma}}
 \simeq\frac{\rho\cos(\phi)}{\sqrt{z^2+\rho^2}},\quad
\hat{\bm{\gamma}}\!\cdot\!\hat{\bm{\alpha}}
 \simeq-1
\end{align}
for $a$ $\!\ll$ $\!r_0$, and Eqs.~(\ref{eq24.15})--(\ref{eq24.17})
[recall Eq.~(\ref{eq24.10})] simplify to 
\begin{align}
\label{B6}
&f_1(\alpha,\beta,\gamma)\simeq\frac{13}{8},\quad
f_2(\alpha,\beta,\gamma)=f_2(\gamma,\alpha,\beta)\simeq\frac{15}{8},
 \nonumber\\
&\qquad f_2(\beta,\gamma,\alpha)\simeq-\frac{3}{4},\quad 
f_3(\alpha,\beta,\gamma)\simeq-\frac{51}{8}.
\end{align}
Substituting Eqs.~(\ref{B2}) and (\ref{B6}) into Eq.~(\ref{B1}),
carrying out the $\phi$-integral, and using 
\begin{equation}
\label{B7}
\int_{-\lambda a}^{\lambda a}\mathrm{d}z\int_0^a\mathrm{d}\rho\,\rho
 \frac{2z^2-\rho^2}{\sqrt{z^2+\rho^2}^5}
 =\frac{2\lambda}{\sqrt{1+\lambda^2}}\,,
\end{equation}
one may find
\begin{equation}
\label{B8}
\Delta_2^{2,\mathrm{c}}U_A(\vec{r}_A)
  =\frac{7\hbar cV\alpha_A(0)\chi^2(0)}
  {256\pi^3\varepsilon_0\rho_A^7}
  \times\frac{\lambda}{\sqrt{1+\lambda^2}}.
\end{equation}
For the integral in $\Delta_2^{2,\mathrm{r}}U_A(\vec{r}_A)$, the
approximations
\begin{align}
\label{B9}
&\alpha=\gamma\simeq\rho_A,\quad
 \beta\simeq 2r_0|\sin(\theta/2)|,\nonumber\\
&\hspace{2ex}\hat{\bm{\alpha}}\!\cdot\!\hat{\bm{\beta}}
 =\hat{\bm{\beta}}\!\cdot\!\hat{\bm{\gamma}}
 \simeq -\frac{r_0|\sin(\theta/2)|}{\rho_A},\nonumber\\
&\hspace{5ex}\hat{\bm{\gamma}}\!\cdot\!\hat{\bm{\alpha}}
 \simeq\frac{2r_0^2\sin^2(\theta/2)}{\rho_A^2}-1
\end{align}
are valid for $a$ $\!\ll$ $\!r_0$. Inspection of Eq.~(\ref{B1})
shows that the leading term in $(a/r_0)$ of
$\Delta_2^{2,\mathrm{r}}U_A(\vec{r}_A)$ is due to the factor
$\beta^{3}$ $\!\propto$ $\sin^{3}(\theta/2)$ in the denominator of the
integrand [cf. Eq.~(\ref{B16}) below], and comes from regions where
$\sin(\theta/2)$ $\!\ll$ $\!1$. Hence we may apply a Taylor
expansion in powers of $\sin(\theta/2)$, retaining only
\begin{equation}
\label{B13}
f_1(\alpha,\beta,\gamma)\simeq\frac{13}{8},\quad
f_2(\beta,\gamma,\alpha)\simeq-\frac{3}{4}.
\end{equation}
Substituting Eqs.~(\ref{B9}) and (\ref{B13}) into Eq.~(\ref{B1}), and
performing the $\theta$-integral using
\begin{equation}
\label{B16}
\int_{\lambda a/r_0}^{2\pi-\lambda a/r_0}\!\!
 \frac{\mathrm{d}\theta}{|\sin^3(\theta/2)|}
 =8\Bigl(\frac{r_0}{\lambda a}\Bigr)^2
 +o\bigl[\ln(\lambda a/r_0)\bigr],
\end{equation}
eventually leads to
\begin{equation}
\label{B17}
\Delta_2^{2,\mathrm{r}}U_A(\vec{r}_A)
  =\frac{7\hbar cV\alpha_A(0)\chi^2(0)}
  {512\pi^3\varepsilon_0\rho_A^7}
  \times\frac{1}{\lambda^2},
\end{equation}
so that
\begin{equation}
\label{B18}
\Delta_2^2U_A(\vec{r}_A)
  =\frac{7\hbar cV\alpha_A(0)\chi^2(0)}
  {512\pi^3\varepsilon_0\rho_A^7}
  \times f(\lambda),
\end{equation}
where
\begin{equation}
\label{B19}
f(\lambda)=\frac{1}{\lambda^2}+\frac{2\lambda}{\sqrt{1+\lambda^2}}
\end{equation}
[recall Eq.~(\ref{B8})]. Note that the approximations made for
calculating $\Delta_2^{2,\mathrm{c}}U_A(\vec{r}_A)$ break down for
large $\lambda$ while those made for calculating
$\Delta_2^{2,\mathrm{r}}U_A(\vec{r }_A)$ break down for small
$\lambda$. We put
\begin{eqnarray}
\label{B20}
f(\lambda)&\mapsto&
{\textstyle\frac{1}{2}}\biggl[
\max_{0.5\le \lambda\le 1.5}f(\lambda)
+\min_{0.5\le \lambda\le 1.5}f(\lambda)
\biggr]\nonumber\\
&&\pm{\textstyle\frac{1}{2}}\biggl[
\max_{0.5\le \lambda\le 1.5}f(\lambda)
-\min_{0.5\le \lambda\le 1.5}f(\lambda)
\biggr]\nonumber\\
&=&3.5\pm 1.4,
\end{eqnarray}
in Eq.~(\ref{B18}), resulting in Eq.~(\ref{eq27.10}).

A similar procedure may be applied in the nonretarded limit, where
Eq.~(\ref{eq24.19}) leads to
\begin{align}
\label{B21}
&\Delta_2^2U_A(\vec{r}_A)
=\frac{3\hbar V}{128\pi^4\varepsilon_0}
 \int_0^{\infty} \mathrm{d} u\, 
 \alpha_A(iu)\chi^2(iu)\nonumber\\
&\quad\times\Biggl\{
 \int_{-\lambda a}^{\lambda a}\mathrm{d}z\int_0^a\mathrm{d}\rho\,\rho
 \int_0^{2\pi}\mathrm{d}\phi
 +\pi a^2\int_{\lambda a/r_0}^{2\pi-\lambda a/r_0}r_0\mathrm{d}\theta
 \Biggr\}\nonumber\\ 
&\quad\times
 \frac{1-3(\hat{\bm{\alpha}}\!\cdot\!\hat{\bm{\beta}})
 (\hat{\bm{\beta}}\!\cdot\!\hat{\bm{\gamma}})
 (\hat{\bm{\gamma}}\!\cdot\!\hat{\bm{\alpha}})}
 {\alpha^3\beta^3\gamma^3}\nonumber\\
&\equiv\Delta_2^{2,\mathrm{c}}U_A(\vec{r}_A)+
 \Delta_2^{2,\mathrm{r}}U_A(\vec{r}_A). 
\end{align}
Use of Eqs.~(\ref{B2}) and (\ref{B7}) leads to
\begin{align}
\label{B22}
\Delta_2^{2,\mathrm{c}}U_A(\rho_A)
=&\,\frac{3\hbar V}
 {64\pi^3\varepsilon_0\rho_A^6}
 \int_0^{\infty} \mathrm{d} u\, \alpha_A(iu)\chi^2(iu)\nonumber\\
 &\,\times\frac{\lambda}{\sqrt{1+\lambda^2}}\,,
\end{align}
while using Eq.~(\ref{B9}), neglecting the term
$(\hat{\bm{\alpha}}\cdot\hat{\bm{\beta}})
(\hat{\bm{\beta}}\cdot\hat{\bm{\gamma}})
(\hat{\bm{\gamma}}\cdot\hat{\bm{\alpha}})$, and recalling
Eq.~(\ref{B16}), results in
\begin{equation}
\label{B23}
\Delta_2^{2,\mathrm{r}}U_A(\rho_A)
=\frac{3\hbar V}
 {128\pi^3\varepsilon_0\rho_A^6}
 \int_0^{\infty} \mathrm{d} u\, \alpha_A(iu)\chi^2(iu)
 \times\frac{1}{\lambda^2}\,.
\end{equation}
Combining Eqs.~(\ref{B22}) and (\ref{B23}) in accordance with
\linebreak Eq.~(\ref{B21}), we obtain
\begin{equation}
\label{B24}
\Delta_2^2U_A(\rho_A)
=\frac{3\hbar V}
 {128\pi^3\varepsilon_0\rho_A^6}
 \int_0^{\infty} \mathrm{d} u\, \alpha_A(iu)\chi^2(iu)
 \times f(\lambda)\,,
\end{equation}
which, in combination with Eq.~(\ref{B20}), implies
Eq.~(\ref{eq27.11}). 


\section{Asymptotic power laws in the case of a half space with
oscillating susceptibility [Eqs.~(\ref{eq13.15}) and (\ref{eq13.16})]}
\label{AppA}

As a preparing step, we derive the linear and quadratic expansions in
$\chi$ of the coefficients 
\begin{align}
\label{eq13.12}
C_4
=&\,\frac{3\hbar c\alpha_A(0)}{64\pi^2\varepsilon_0}
 \int_{1}^\infty\mathrm{d} v\,
 \Biggl\{\frac{[\chi(0)+1]v-\sqrt{\chi(0)+v^2}}
 {[\chi(0)+1]v+\sqrt{\chi(0)+v^2}}
 \nonumber\\
&\,\times\!\biggl(\frac{2}{v^2}-\frac{1}{v^4}\biggr)
 -\frac{v-\sqrt{\chi(0)+v^2}}{v+\sqrt{\chi(0)+v^2}}
 \times\frac{1}{v^4}\Biggr\}\nonumber\\
=&\,\Delta_1C_4+\Delta_2C_4+\ldots
\end{align}
and
\begin{align}
\label{eq13.14}
C_3
=&\,\frac{\hbar}{16\pi^2\varepsilon_0}
 \int_0^\infty\mathrm{d}u\ \alpha_A(iu)
 \frac{\chi(iu)}{\chi(iu)+2}\nonumber\\
=&\,\Delta_1C_3+\Delta_2C_3+\ldots\,,
\end{align}
that can be found for the retarded and nonretarded distance laws of
the homogeneous half space \cite{Buhmann05}. Substituting
\begin{align}
\label{A1}
&\frac{[\chi(0)+1]v-\sqrt{\chi(0)+v^2}}
 {[\chi(0)+1]v+\sqrt{\chi(0)+v^2}}\nonumber\\
&\quad=\biggl[\frac{1}{2}-\frac{1}{4v^2}\biggr]\chi(0)
 -\biggl[\frac{1}{4}-\frac{1}{8v^2}\biggr]\chi^2(0)
 +\ldots\,,\\
\label{A2}
&\frac{v-\sqrt{\chi(0)+v^2}}{v+\sqrt{\chi(0)+v^2}}
=-\frac{1}{4v^2}\chi(0)+\frac{1}{8v^4}\chi^2(0)+\ldots
\end{align}
into Eq.~(\ref{eq13.12}) and carrying out the remaining $v$-integral,
we arrive at Eqs.~(\ref{A3}) and (\ref{A4}), while Eq.~(\ref{eq13.14})
implies Eqs.~(\ref{A5}) and (\ref{A6}).

The spatial integrals in Eqs.~(\ref{eq13.4}) [recall
Eq.~(\ref{eq13.5})], (\ref{eq13.8})  and (\ref{eq13.9}) can be carried
out explicitly for $p(z)$ $\!=$ $\!\cos^2(k_zz)$,
\begin{align}
\label{A7}
&\int_0^\infty\mathrm{d}z\,e^{-2bz}\cos^2(k_zz)
=\frac{2b^2+k_z^2}{4b(b^2+k_z^2)}\,,\\
\label{A8}
&\int_0^\infty\mathrm{d}z\,\cos^2(k_zz)
\int_z^\infty\mathrm{d}z'\,e^{-2bz'}\cos^2(k_zz')\nonumber\\
&\quad=\frac{2b^6+8b^4k_z^2+5b^2k_z^4+2k_z^6}
 {8b^2(b^2+k_z^2)^2(b^2+4k_z^2)}\,,
\end{align}
resulting in
\begin{align}
\label{A9}
&\Delta_1^1U_A(z_A)=-\frac{\hbar\mu_0}{16\pi^2}
 \int_0^{\infty}\!\!\mathrm{d}u\,u^2 \alpha_A(iu)\chi(iu)
 \int_0^\infty\!\!\mathrm{d}q\,\frac{q}{b}
 \nonumber\\
&\quad\times e^{-2bz_\mathrm{A}}\frac{2b^2+k_z^2}{b^2+k_z^2}
 \Biggl[\biggl(\frac{bc}{u}\biggr)^{\!\!2}\hspace{-0.5ex}-1
 +\frac{1}{2}\biggl(\frac{u}{bc}\biggr)^{\!\!2}\Biggr],\\
\label{A10}
&\Delta_2^1U_A(z_A)=\frac{\hbar\mu_0}{16\pi^2}
 \!\int_0^{\infty}\!\!\mathrm{d}u\,u^2 \alpha_A(iu)\chi^2(iu)
 \!\int_0^\infty\!\!\mathrm{d}q\,\frac{q}{b}\nonumber\\
&\quad\times e^{-2bz_\mathrm{A}}\frac{2b^2+k_z^2}{b^2+k_z^2}
\Biggl[\frac{1}{2}\biggl(\frac{bc}{u}\biggr)^{\!\!2}
 \hspace{-0.5ex}-\frac{3}{4}
 +\frac{1}{4}\biggl(\frac{u}{bc}\biggr)^{\!\!2}
 \Biggr],\\
\label{A11}
&\Delta_2^2U_A(z_A)=\frac{\hbar\mu_0}{16\pi^2}
 \!\int_0^{\infty}\!\!\mathrm{d}u\,u^2 \alpha_A(iu)\chi^2(iu)
 \!\int_0^\infty\!\!\mathrm{d}q\,\frac{q}{b}\nonumber\\
&\quad\times e^{-2bz_\mathrm{A}}
 \frac{2b^6+8b^4k_z^2+5b^2k_z^4+2k_z^6}
 {(b^2+k_z^2)^2(b^2+4k_z^2)}
 \nonumber\\
&\quad\times\Biggl[\frac{1}{2}
 -\frac{1}{2}\biggl(\frac{u}{bc}\biggr)^{\!\!2}
 \hspace{-0.5ex}+\frac{1}{4}\biggl(\frac{u}{bc}\biggr)^{\!\!4}
 \Biggr].
\end{align}

In analogy to the procedure outlined in Ref.~\cite{Buhmann05}, the
retarded limit may conveniently be treated by introducing the new
integration variable $v$ $\!=$ $\!bc/u$, transforming integrals
according to
\begin{align}
\label{A13}
\int_0^\infty\mathrm{d}u\,u^2 &
 \int_0^\infty\mathrm{d}q\,\frac{q}{b}\,e^{-2bz_\mathrm{A}}\ldots
 \nonumber\\
&\mapsto c^3\int_1^\infty\frac{\mathrm{d}v}{v^4}
 \int_0^\infty\mathrm{d}b\,
 b^3\,e^{-2bz_\mathrm{A}}\ldots\ ,
\end{align}
applying the approximation (\ref{eq20}), and carrying out the
$v$-integrals. Application of this procedure to Eqs.~(\ref{A9}),
(\ref{A10}), and (\ref{A11}) leads to
\begin{align}
\label{A14}
&\Delta_1^1U_A(z_A)
 =-\frac{23\hbar c\alpha_A(0)\chi(0)F_3(k_zz_A)}
 {640\pi^2\varepsilon_0z_A^4}\,,\\
\label{A15}
&\Delta_2^1U_A(z_A)
 =\frac{9\hbar c\alpha_A(0)\chi^2(0)F_3(k_zz_A)}
 {640\pi^2\varepsilon_0z_A^4}\,,\\
\label{A16}
&\Delta_2^2U_A(z_A)
 =\frac{43\hbar c\alpha_A(0)\chi^2(0)H_3(k_zz_A)}
 {8960\pi^2\varepsilon_0z_A^4}\,,
\end{align}
where we have introduced the definitions (\ref{eq13.17}) and
(\ref{eq13.18}). Combining Eqs.~(\ref{A14})--(\ref{A16}) in
accordance with Eq.~(\ref{eq13}) and using Eqs.~(\ref{A3}) and
(\ref{A4}), we arrive at Eq.~(\ref{eq13.15}).

The asymptotic behavior of Eqs.~(\ref{A9})--(\ref{A11}) in the
nonretarded limit may be obtained by transforming the integral
according to
\begin{align}
\label{A17}
\int_0^\infty\mathrm{d}u &
 \int_0^\infty\mathrm{d}q\,\frac{q}{b}\,e^{-2bz_\mathrm{A}}\ldots
\nonumber\\
&\mapsto\int_0^\infty\mathrm{d}u\,
 \int_0^\infty\mathrm{d}b\,
 e^{-2bz_\mathrm{A}}\ldots\ ,
\end{align}
retaining only the leading power of $u/(bc)$, carrying out the
$b$-integral and discarding higher-order terms in $uz_\mathrm{A}/c$
(cf. Ref.~\cite{Buhmann05}), resulting in
\begin{align}
\label{A18}
&\Delta_1^1U_A(z_A)
 =-\frac{\hbar F_2(k_zz_A)}
 {32\pi^2\varepsilon_0z_A^3}
 \int_0^\infty\mathrm{d}u\,\alpha_A(iu)\chi(iu)\,,\\
\label{A19}
&\Delta_2^1U_A(z_A)
 =\frac{\hbar F_2(k_zz_A)}
 {64\pi^2\varepsilon_0z_A^3}
 \int_0^\infty\mathrm{d}u\,\alpha_A(iu)\chi^2(iu)\,,\\
\label{A20}
&\Delta_2^2U_A(z_A)
 =\frac{\hbar\mu_0 H_0(k_zz_A)}
 {32\pi^2z_A}
 \int_0^\infty\mathrm{d}u\,u^2\alpha_A(iu)\chi^2(iu)\,,
\end{align}
recall Eqs.~(\ref{eq13.17}) and (\ref{eq13.18}). Upon using
Eq.~(\ref{eq13}), Eqs.~(\ref{A18}) --(\ref{A20}) lead to
Eq.~(\ref{eq13.16}), where we have neglected the term proportional to
$z_A^{-1}$ in consistency with the nonretarded limit and used
Eqs.~(\ref{A5}) and (\ref{A6}).



\begin{thebibliography}{99}

\bibitem{Dzyaloshinskii61}
I. E. Dzyaloshinskii, E. M. Lifshitz, and L. P. Pitaevskii: Adv. Phys.
\textbf{10}, 165 (1961).

\bibitem{Langbein74}
D. Langbein: Springer Tracts Mod. Phys.
\textbf{72}, 1 (1974).

\bibitem{Mahanty76}
J. Mahanty and B. W. Ninham: \textit{Dispersion
Forces} (Academic, London, 1976).

\bibitem{Hinds91}
E. A. Hinds: in \textit{Advances in
Atomic, Molecular, and Optical Physics}, edited by D. Bates and B.
Bederson (Academic, New York, 1991), Vol. 28, p. 237.

\bibitem{Milonni94}
P. W. Milonni: \textit{The Quantum Vacuum: An Introduction to Quantum
Electrodynamics} (Academic, San Diego, 1994).

\bibitem{Casimir48}
H. B. G. Casimir and D. Polder: Phys. Rev. \textbf{73}, 360 (1948).

\bibitem{Casimir48b}
H. B. G. Casimir: Proc. K. Ned. Akad. Wet. \textbf{51}, 793 (1948).

\bibitem{Nelson02}
D. L. Nelson and M. M. Cox: \textit{Lehninger Principles of
Biochemistry} (Worth Publishers, New York, 2002), pp. 177, 250--253,
331, 365, 1053.

\bibitem{Autumn02}
K. Autumn, M. Sitti, Y. A. Liang, A. M. Peattie, W. R. Hansen, S.
Sponberg, T. W. Kelly, R. Fearing, J. N. Israelachvili, and R. J.
Full: PNAS \textbf{99}, 19, 12252 (2002).

\bibitem{Kesel04}
A. B. Kesel, A. Martin, and T. Seidl: Smart Mater. Struct.
\textbf{13}, 512 (2004).

\bibitem{Binnig86}
G. Binnig, C. Gerber, and C. F. Quate: Phys. Rev. Lett. \textbf{56},
9, 930 (1986); for a review see F. J. Giessibl: Rev. Mod. Phys.
\textbf{75}, 949 (2003).

\bibitem{Henkel04}
C. Henkel and K. Joulain: e-print \texttt{quant-ph/0407153}.

\bibitem{Bruch83}
L. W. Bruch: Surf. Sc. \textbf{125}, 194 (1983).

\bibitem{Shimizu02}
V. I. Balykin, V. S. Letokhov, Y. B. Ovchinnikov, and A. I. Sidorov:
Phys. Rev. Lett. \textbf{60}, 21, 2137 (1988); F. Shimizu and J.-i.
Fujita: Phys. Rev. Lett. \textbf{88}, 12, 123201 (2002).

\bibitem{Lin04}
Y.-y. Lin, I. Teper, C. Chin, and V. Vuleti\'{c}: Phys. Rev. Lett.
\textbf{92}, 5, 050404 (2004).

\bibitem{McLachlan63}
A. D. McLachlan: Proc. R. Soc. London Ser. A \textbf{271}, 387 (1963).

\bibitem{McLachlan63b}
A. D. McLachlan: Mol. Phys. \textbf{7}, 381 (1963).

\bibitem{Tiko93}
Y. Tikochinsky and L. Spruch: Phys. Rev. A \textbf{48}, 6, 4223
(1993).

\bibitem{Enderlein99}
S.-T. Wu and C. Eberlein: Proc. R. Soc. London Ser. A \textbf{455},
2487 (1999).

\bibitem{Zhou95}
F. Zhou and L. Spruch: Phys. Rev. A \textbf{52}, 297 (1995). 

\bibitem{Wylie84}
J. M. Wylie and J. E. Sipe: Phys. Rev. A \textbf{30}, 1185 (1984).

\bibitem{Jhe91}
W. Jhe: Phys. Rev. A \textbf{43}, 11, 5795 (1991). 

\bibitem{Buhmann05}
S. Y. Buhmann, Ho Trung Dung, T. Kampf, and D.-G. Welsch: Eur. Phys.
J. D \textbf{35}, 15 (2005).

\bibitem{Buhmann05b}
S. Y. Buhmann, T. Kampf, and D.-G. Welsch: Phys. Rev. A \textbf{72},
032112 (2005).

\bibitem{Marvin82}
A. M. Marvin and F. Toigo: Phys. Rev. A \textbf{25}, 2, 782 (1982).

\bibitem{Girard89}
C. Girard, S. Maghezzi, and F. Hache: J. Chem. Phys. \textbf{91}, 5509
(1989).

\bibitem{Boustimi02}
M. Boustimi, J. Baudon, P. Candori, and J. Robert: Phys. Rev. B
\textbf{65}, 155402 (2002).

\bibitem{Buhmann04}
S. Y. Buhmann, Ho Trung Dung, and D.-G. Welsch: J. Opt. B: Quantum
Semiclass. Opt. \textbf{6}, 127 (2004).

\bibitem{Buhmann04b}
S. Y. Buhmann, Ho Trung Dung, L. Kn\"{o}ll, and D.-G. Welsch:
Phys. Rev. A \textbf{70}, 052117 (2004).

\bibitem{Lifshitz56}
E. M. Lifshitz: Soviet Phys. JETP \textbf{2}, 1, 73 (1956).

\bibitem{Parsegian74}
V. A. Parsegian: Mol. Phys. \textbf{27}, 6, 1503 (1974).

\bibitem{Schwinger78}
J. Schwinger, L.L. DeRaad, Jr., and K. A. Milton: Ann. Phys. (N. Y.)
\textbf{115}, 1, 1, (1978).

\bibitem{Mil92}
P. W. Milonni and M.-L. Shih: Phys. Rev. A \textbf{45}, 7, 4241
(1992).

\bibitem{Kup92}
D. Kupiszewska: Phys. Rev. A \textbf{46}, 5, 2286 (1992).

\bibitem{Barton01}
G. Barton: J. Phys. A \textbf{34}, 4083 (2001).

\bibitem{Raa05}
C. Raabe and D.-G. Welsch: Phys. Rev. A  \textbf{71}, 013814 (2005).

\bibitem{Renne67}
B. R. A. Nijboer and M. J. Renne: Chem. Phys. Lett. \textbf{1}, 317
(1967); M. J. Renne: Physica \textbf{56}, 193 (1970); \textit{ibid.}
\textbf{56}, 125 (1971).

\bibitem{Milonni92b}
P. W. Milonni and P. B. Lerner: Phys. Rev. A \textbf{46}, 3, 1185
(1992).

\bibitem{Axilrod43}
B. M. Axilrod and E. Teller: J. Chem. Phys. \textbf{11}, 6, 299
(1943); B. M. Axilrod: J. Chem. Phys. \textbf{17}, 1349 (1949);
\textit{ibid.} \textbf{19}, 6, 719 (1951).

\bibitem{Aub60}
M. R. Aub and S. Zienau: Proc. R. Soc. London, Ser. A \textbf{257},
464 (1960). 

\bibitem{Power85}
E. A. Power and T. Thirunamachandran: Proc. R. Soc. London Ser. A
\textbf{401}, 267 (1985).

\bibitem{Power94}
E. A. Power and T. Thirunamachandran: Phys. Rev. A \textbf{50}, 5,
3929 (1994).

\bibitem{Cirone96}
M. Cirone and R. Passante: J. Phys. B \textbf{29}, 1871 (1996).

\bibitem{Knoll01}
L. Kn\"{o}ll, S. Scheel, and D.-G. Welsch:
in \textit{Coherence and Statistics of Photons and Atoms},
edited by J. Pe\v{r}ina (Wiley, New York, 2001), p. 1;
(for an update, see \texttt{arXiv:quant-ph/0006121}).

\bibitem{Jackson99}
J. D. Jackson: \textit{Classical Electrodynamics} (Wiley, New York,
1999), p. 162.

\bibitem{Safari05}
H. Safari, S. Y. Buhmann, Ho Trung Dung, and D.-G. Welsch: in
preparation for publication.
\end{thebibliography}
\end{document}